\documentclass[12pt]{iopart}
\usepackage{iopams}
\usepackage{graphicx}

\input epsf.sty

\newlength{\TZ}
\newcommand{\BEQ}{\begin{equation}}     
\newcommand{\BEA}{\begin{eqnarray}}
\newcommand{\BD}{\begin{displaymath}}
\newcommand{\EEQ}{\end{equation}}       
\newcommand{\EEA}{\end{eqnarray}}
\newcommand{\ED}{\end{displaymath}}
\newcommand{\eps}{\varepsilon}          
\newcommand{\vph}{\varphi}              
\newcommand{\D}{{\rm d}}                
\newcommand{\II}{{\rm i}}               

\newcommand{\wit}[1]{\widetilde{#1}}    

\renewcommand{\vec}[1]{\boldsymbol{#1}} 

\newcommand{\zeile}[1]{\vskip #1 \baselineskip} 

\newcommand{\half}{{1\over 2}}

\newcommand{\C}{\mathbb{C}}

\newcommand{\annexe}[1]{\setcounter{equation}{0}\setcounter{subsection}{0}
\section*{Appendix. #1}
\renewcommand{\theequation}{A\arabic{equation}}
              \renewcommand{\thesection}{A} }

\catcode`\@=11
\def\numberbysection{\@addtoreset{equation}{section}
        \def\theequation{\thesection.\arabic{equation}}}

\begin{document}


\title[Autocorrelation functions]{Autocorrelation functions in 
phase-ordering kinetics from local scale-invariance}

\author{Malte Henkel$^{1}$ and Florian Baumann$^{1,2}$}
\address{$^1$Laboratoire de Physique des 
Mat\'eriaux,\footnote{Laboratoire associ\'e au CNRS UMR 7556} 
Universit\'e Henri Poincar\'e Nancy I, \\ 
B.P. 239, F -- 54506 Vand{\oe}uvre l\`es Nancy Cedex, France}
\address{$^2$Institut f\"ur Theoretische Physik I, 
Universit\"at Erlangen-N\"urnberg, \\
Staudtstra{\ss}e 7B3, D -- 91058 Erlangen, Germany}

\begin{abstract}
The explicit calculation of the scaling form of the two-time
autocorrelation function in phase-ordering kinetics and in those
cases of non-equilibrium critical dynamics where the dynamical exponent
$z=2$ through the extension of dynamical scaling to local scale-invariance 
is reviewed. Conceptually, this mainly 
requires an extension from the usually considered
$d$-dimensional ageing or Schr\"odinger algebras to a new kind of representation 
of the conformal algebra in $d+2$ dimensions. Explicit tests in several 
exactly solved models of simple magnets and through simulations in the
$2D$ Ising and $q$-states Potts models ($q=2,3,8$) quenched to $T<T_c$ 
are presented and the
extension to systems with non-equilibrium steady-states is discussed through
two exactly solvable models as well. In the context of surface growth models,
possible generalizations for a dynamical exponent $z=4$ and beyond are
discussed. 
 
\end{abstract}

\submitto{J. Stat. Mech.}
\maketitle

\setcounter{footnote}{0}

\section{Introduction}

Phase-ordering kinetics \cite{Bray94} 
may occur if a magnet is rapidly brought (`quenched') 
{}from some initial disordered state to some temperature $T<T_c$ below
its critical temperature $T_c>0$. At equilibrium, such systems have at least 
two equivalent macroscopically ordered thermodynamic states. Because of the 
competition between these two equilibrium states, the system cannot
simply relax within a finite time towards one of them but rather, stimulated by
the small effective magnetic fields which are created by the environment of
each magnetic atom, ordered domains will form and subsequently grow. Although
the average magnetization, if initially zero, will {\em not} change, there is
a slow, non-exponential relaxation as the walls separating the ordered domains
slowly move. For a spatially infinite system, this gradual coarsening process
will continue indefinitely, and the state of the system and the behaviour
of time-dependent observables will depend on the age of the system, that is
the time elapsed since the quench. In addition, as will be discussed in
detail in this article, dynamical scaling is observed, although neither of the
equilibrium states by itself is scale-invariant. The three properties
(i) slow, non-exponential dynamics, (ii) breaking of time-translation 
invariance and (iii) some kind of dynamical scaling are the constitutive
properties of {\em ageing systems}.\footnote{In distinction to
chemical/biological ageing, `physical ageing' as defined here can be observed
even if the underlying microscopic dynamics is completely reversible.}  

Physical ageing was  first identified in celebrated experiments on the 
mechanical response of certain glassy systems to external stress \cite{Stru78}. 
For a better theoretical understanding, the quite recent realization that 
very similar phenomena can also be found in simple magnets, without disorder 
nor frustrations, might lead
to conceptual insights which in turn could become also fruitful in more
complex systems. For recent reviews of this intensively studied topic, see
\cite{Cugl02,Godr02,Cris03,Cala05,Cham05,Gamb06,Henk06,Henk07a,Henk07b}. 

To begin, we recall some facts about dynamical scaling of phase-ordering 
systems. Unless explicitly stated otherwise, we assume 
that throughout the order-parameter is {\em non-conserved} by the dynamics and 
that the initial state is totally disordered. For late time, 
linear size $L$ of the ordered domains grows as $L\sim t^{1/z}$ and it is
known that the dynamical
exponent $z=2$ for a non-conserved order-parameter \cite{Bray94b}. 
If $\phi(t,\vec{r})$ is the order-parameter at time $t$ and location
$\vec{r}$ the following dynamical scaling of
the two-time autocorrelation $C$ and (linear) autoresponse function $R$ is
usually assumed 
\BEA
C(t,s) &=& \langle \phi(t,\vec{r}) \phi(s,\vec{r}) \rangle ~~ 
\sim s^{-b} f_C(t/s) 
\label{gl:sC} \\
R(t,s) &=& \left.\frac{\delta\langle\phi(t,\vec{r})\rangle}{\delta 
h(s,\vec{r})}\right|_{h=0}
\sim s^{-1-a} f_{R}(t/s)
\label{gl:sR}
\EEA
where $h(s,\vec{r})$ is the conjugate
magnetic field at time $s$ and location $\vec{r}$. 
The scaling behaviour is expected to apply
in the so-called {\em ageing regime} where 
\BEQ \label{1:validite}
t,s\gg t_{\rm micro} \mbox{\rm ~~and~~} t-s\gg t_{\rm micro}
\EEQ
where $t_{\rm micro}$ is a microscopic time-scale \cite{Zipp00}. In writing 
eqs.~(\ref{gl:sC},\ref{gl:sR}) 
it was tacitly assumed that the scaling derives from the 
algebraic time-dependence of the single characteristic length-scale 
$L=L(t)\sim t^{1/z}$ which measures the linear size of 
correlated or ordered clusters and where $z$ is the dynamic exponent. 
Under these assumptions, the above forms define the 
non-equilibrium exponents $a$ and $b$ and the scaling functions $f_C(y)$ and
$f_R(y)$. For large arguments $y=t/s\to \infty$, one generically expects
\BEQ \label{1:CR}
f_C(y) \sim y^{-\lambda_C/z} \;\; , \;\;
f_R(y) \sim y^{-\lambda_R/z}
\EEQ
where $\lambda_C$ and $\lambda_R$, respectively, are known as autocorrelation
\cite{Fish88,Huse89} and autoresponse exponents \cite{Pico02}. 
In non-disordered magnets with short-ranged initial conditions one usually
has $\lambda_C=\lambda_R$, but this is need no longer be true if either
of these conditions is relaxed \cite{Newm90,Pico02,Sche03,Henk06b}. 
Field-theoretical considerations for systems quenched to $T=T_c$ 
show that for a non-conserved order-parameter 
the calculation of $\lambda_{C,R}$ requires an independent renormalization 
and hence one cannot expect to find a scaling relation between
these and equilibrium exponents (including $z$) 
\cite{Jans89}.\footnote{That is
different if a non-vanishing initial value of the order-parameter is 
considered \cite{Cala06,Cala07,Fedo06,Anni06,Baum06d}.} For phase-ordering 
(quenches to $T<T_c$) a well-known result due to Bray \cite{Bray94} 
relates the difference $\lambda_C-\lambda_R$ to an exponent describing the
fall-off of the initial correlator. In particular  it can be shown 
that $\lambda_C=\lambda_R$ for short-ranged initial correlations \cite{Bray94}. 
This same result can alternatively be derived from local scale-invariance 
\cite{Pico04} (see section~3). 

The values of the exponents $a$ and $b$ are as follows for phase-ordering. 
First, one usually observes simple scaling of $C(t,s)=f_C(t/s)$, 
hence $b=0$.\footnote{Explicit results on the ageing behaviour of the 
spherical model close to free surface give $b\ne 0$ \cite{Baum06a}.} 
Second, the value of $a$ depends on whether the equilibrium
correlator decays exponentially (`short-ranged') or algebraically 
(`long-ranged'). This defines the 
classes S and L, respectively and one has \cite{Bert99,Cate00,Henk02a,Henk03e}
\BEQ
\hspace{-1truecm} C_{\rm eq}(\vec{r})\sim \left\{\begin{array}{l} 
e^{-|\vec{r}|/\xi} \\
|\vec{r}|^{-(d-2+\eta)} \end{array} \right. 
\;\; \Longrightarrow \;\; 
\left\{\begin{array}{l} \mbox{\rm class S} \\ 
\mbox{\rm class L}\end{array}\right.
\;\; \Longrightarrow \;\; 
a = \left\{ \begin{array}{c} 1/z \\ (d-2+\eta)/z \end{array} \right.
\EEQ
Examples for short-ranged models (class S) include the Ising or Potts 
models in $d>1$ dimensions, while 
the spherical model or the $2D$ XY model below the Kosterlitz-Thouless 
transition are examples for long-ranged systems 
(class L).\footnote{For quenches to $T=T_c$, all systems are of course in 
class L.}  

A much harder problem is posed by the quest to find the universal 
scaling functions $f_R(y)$ and $f_C(y)$. Indeed, 
one might consider the analogy with 
equilibrium critical phenomena where simple scale-invariance can be
extended to {\em conformal} invariance, under quite general conditions. 
Since conformal transformations are scale-transformations with a
spatially varying rescaling factor (such that angles are conserved), 
one can inquire about the possibility to similarly extend dynamical scaling to
an invariance of the physical system under more general, {\em local} 
scale-transformations \cite{Henk94,Henk02}. 
Specifically, motivated by the analogy with conformal
invariance \cite{Poly70,Bela84}, the transformations of time
\BEQ \label{1:tt}
t \mapsto t' = \frac{\alpha t +\beta}{\gamma t +\delta} \;\; ; \;\;
\alpha\delta-\beta\gamma=1
\EEQ
were considered to which transformations of space $\vec{r}\mapsto \vec{r}'$
must be added in such a way that,say, these local scale-transformation form 
a Lie algebra. Furthermore, it can be argued that the linear response functions
should transform covariantly under such local scale-transformations (at least 
if the scaling operators from which they are built are so-called 
{\em quasi-primary} operators) \cite{Henk94,Henk02,Pico04}. Taking 
into account
that time-translations cannot be part of local scale-transformations as
applied to ageing phenomena (hence $\beta=0$ in (\ref{1:tt})), 
this leads to \cite{Pico04,Henk06a,Henk03} ($\Theta$ is the Heaviside function)
\BEQ \label{1:R}
f_R(y) = f_0 \, y^{1+a'-\lambda_R/z} (y-1)^{-1-a'} \Theta(y-1)
\EEQ
such that the form of $f_R(y)$ is determined by the two exponents
$\lambda_R/z$ and $a'$ which are related to the scaling dimensions of the
order-parameter $\phi$ and the associate response field $\wit{\phi}$ 
and $f_0$ is a normalization constant. By now there exists quite a long list of
models where either the exact solution or numerical data for 
$R(t,s)=\langle \phi(t)\wit{\phi}(s)\rangle$ are 
compatible with the prediction (\ref{1:R}) of local scale-invariance as 
reviewed in detail in \cite{Henk07a,Henk07b}. 

Here, we shall address the question how to find $f_C(y)$ from an extension
of dynamical scaling to local scale-invariance. Although the method
has  been successfully applied in various contexts, 
see \cite{Henk04b,Baum05b,Lore07,Jank06,Henk06a,Roet06,Baum07}, only
the barest outline of it was ever published \cite{Henk04b,Henk06a}. 
As we shall see, the analytical calculation of correlators is quite 
a demanding task, be it in the context of closed-form approximations 
\cite{Bray91,Toyo92,Roja99}, perturbative schemes \cite{Liu91,Maze98,Maze04} 
or else from a dynamical symmetry. Since $z=2$ in phase-ordering kinetics, 
a natural candidate for a local scale-symmetry is
provided by the well-known Schr\"odinger group and/or some of its subgroups. 
We shall recall their definition in section~2 and remind that the
Schr\"odinger group is a group of dynamical symmetries of the free diffusion 
equation. It may be less well-known that, provided one considers the
diffusion constant as an additional variable, there is a larger dynamical
symmetry isomorphic to a conformal group in $d+2$ 
dimensions \cite{Burd73,Henk03}. These results
may be extended to slightly more general diffusion equations. Next, we consider
the Langevin description for phase-ordering kinetics where there exists 
an exact reduction formula which reduces the calculation
of any $n$-point correlation/response function to the calculation of 
certain correlators/responses in the {\em noiseless} part of the theory only,
given that so-called Bargman superselection rules for that deterministic
part are valid. In section~3, we recall first the standard result from 
Schr\"odinger-invariance
for the required three-point function and then show how using the previously
discussed extension to a conformal group the two-time correlation function can
be found. In section~4 we review existing tests of this prediction, both
analytical and numerical. Finally, in section~5 we present a short outlook
how the methods reviewed here might be extendable to cases where $z\ne 2$
and conclude in section~6. A technical point on the determination of the
autocorrelation function is treated in the appendix. 

\section{Dynamical symmetries: Schr\"odinger, ageing and conformal}

In this section, we recall those elements of local scale-invariance, 
specialized to the case $z=2$, which we shall need for the explicit
calculation of two-time correlations. 

\subsection{Schr\"odinger-invariance of the free diffusion equation}

Consider the dynamical symmetries of the free diffusion 
(or free Schr\"odinger) equation
\BEQ \label{gl:diffu}
2{\cal M} \partial_t \phi = \Delta \phi
\EEQ
where $\Delta=\vec{\nabla}\cdot\vec{\nabla}$ is the spatial laplacian and 
the `mass' $\cal M$ plays the r\^ole of a kinetic coefficient. The 
{\em Schr\"odinger-group} {\sl Sch}($d$) 
(actually, already found by Lie more than a century 
ago) contains the space-time transformations
\BEQ \label{gl:5:2:Schr}
t \mapsto t' = \frac{\alpha t+\beta}{\gamma t+\delta} \;\; ; \;\;
\vec{r} \mapsto \vec{r}' = 
\frac{{\cal R}\vec{r} + \vec{v}t + \vec{a}}{\gamma t+\delta} 
\;\; , \;\;
\alpha\delta - \beta\gamma =1
\EEQ
where $\alpha,\beta,\gamma,\delta,\vec{v},\vec{a}$ are real (vector) 
parameters and ${\cal R}$ is a rotation matrix in $d$ spatial dimensions. 
The group acts projectively on a solution $\phi$ of the diffusion 
equation through 
$(t,\vec{r})\mapsto g(t,\vec{r})$, $\phi\mapsto T_g \phi$
\BEQ \label{gl:5:2:Schrpsi}
\left(T_g \phi\right)(t,\vec{r}) = 
f_g(g^{-1}(t,\vec{r}))\,\phi(g^{-1}(t,\vec{r}))
\EEQ
where $g$ is an element of the Schr\"odinger group and the companion 
function reads \cite{Nied72,Perr77}
\BEQ \label{gl:5:2:Schrf}
\hspace{-2.2truecm}
f_{g}(t,\vec{r}) = (\gamma t+\delta)^{-d/2} 
\exp\left[ -\frac{{\cal M}}{2} \frac{\gamma \vec{r}^2+
2{\cal R}\vec{r}\cdot(\gamma\vec{a}-\delta\vec{v})+\gamma\vec{a}^2-
t\delta\vec{v}^2+2\gamma\vec{a}\cdot\vec{v}}{\gamma t+\delta}\right]
\EEQ
It is then natural to include also arbitrary phase-shifts of the wave function 
$\phi$ within the Schr\"odinger group {\sl Sch}($d$). We shall 
denote the Lie algebra of {\sl Sch}($d$) by $\mathfrak{sch}_d$. 
The invariance of the space of solutions of the free diffusion equation 
(\ref{gl:diffu}) 
under this group can be illustrated in $d=1$ by introducing the 
Schr\"odinger operator
\BEQ
{\cal S} := 2 M_0 Y_{-1} - Y_{-1/2}^2 = 2 {\cal M} \partial_t - \partial_r^2
\EEQ
In what follows, we shall restrict calculations 
to the case $d=1$ since the extensions to \\
$d>1$ will be obvious. The Schr\"odinger Lie algebra
$\mathfrak{sch}_1=\langle X_{-1,0,1},Y_{-\frac{1}{2},\frac{1}{2}},M_0\rangle$
is spanned by the infinitesimal generators of temporal and spatial
translations ($X_{-1},Y_{-1/2}$), Galilei-transformations ($Y_{1/2}$),
phase shifts ($M_0$), space-time dilatations with $z=2$ ($X_0$) and so-called
special transformations ($X_1$). The generators read explicitly \cite{Henk94}
\BEA
X_n &=& -t^{n+1}\partial_t -\frac{n+1}{2} t^n r\partial_r -\frac{n(n+1)}{4}
{\cal M} t^{n-1} r^2 - \frac{x}{2}(n+1) t^n \nonumber \\
Y_m &=& -t^{m+1/2}\partial_r -\left( m+\frac{1}{2}\right) {\cal M} t^{m-1/2} r
\label{gl:5:2:SchrGen} \\
M_n &=& -{\cal M} t^n \nonumber
\EEA
Here $x$ is the scaling dimension and ${\cal M}$ is the {mass} of the
scaling operator $\phi$ on which these generators act. 
The non-vanishing commutation relations are
\BEA
\left[ X_n , X_{n'} \right] &=& (n-n') X_{n+n'} \;\; , \;\;
\left[ X_n , Y_m \right] \:=\: \left(\frac{n}{2}-m\right) Y_{n+m} 
\nonumber \\ 
\left[ X_n , M_{n'} \right] &=& -n' M_{n+n'} \;\; , \;\; 
\left[ Y_m , Y_{m'} \right] \:=\: (m-m') M_{m+m'} 
\label{gl:5:2:SchAlg}
\EEA
\setcounter{footnote}{1}
\typeout{*** Hier wird der Fussnotenzaehler zurueckgestellt ***}
The invariance of the $1D$ diffusion equation under the action of 
$\mathfrak{sch}_1$
is now seen from the following commutators which follow from the
explicit form (\ref{gl:5:2:SchrGen})
\BEA
\left[{\cal S}, X_{-1}\right] &=& \left[ {\cal S}, Y_{\pm 1/2} \right]
\:=\: \left[ {\cal S}, M_0\right] \:=\: 0 
\nonumber \\
\left[{\cal S}, X_0 \right] &=& -{\cal S}
\;\; , \;\; \qquad \;\;\:
\left[ {\cal S}, X_1 \right] \:=\: -2t {\cal S} - (2x-1) M_0 
\label{2:Sinv}
\EEA
Therefore, in $d$ spatial dimensions, {\em for any solution $\phi$ 
of the Schr\"odinger equation ${\cal S}\phi=0$ with
scaling dimension $x=d/2$, the infinitesimally transformed solution 
${\cal X}\phi$ with ${\cal X}\in\mathfrak{sch}_d$ also satisfies the 
Schr\"odinger equation ${\cal S}{\cal X}\phi=0$} \cite{Kast68,Nied72,Hage72}. 

Technically, we have been making an important assumption here. Indeed, 
and borrowing terminology from conformal invariance \cite{Bela84}, only
so-called {\em quasiprimary} scaling operators $\phi$ will transform under the
action of the Schr\"odinger group such that the change under an infinitesimal
transformation, generated by 
${\cal X}\in\mathfrak{sch}_d$, is simply given by
${\cal X}\phi$ \cite{Henk94,Henk02}. It is non-trivial that a 
given physical observable
should be represented by a quasiprimary scaling operator, although this 
appears usually to be the case for the order-parameter. 
For example, if $\phi=\phi(t,\vec{r})$ is 
quasiprimary, neither of the derivatives $\partial_t\phi$ nor 
$\partial_{\vec{r}}\phi$ is. 

Now the form of $n$-point functions built from quasiprimary
scaling operators can be constrained by writing down differential
equations ${\cal X}^{[n]}\langle \phi_1 \ldots \phi_n\rangle=0$ 
with the $n$-particle
extensions ${\cal X}^{[n]}$ of the generators ${\cal X}\in\mathfrak{sch}_d$.  
Explicit results will be quoted below.  

\subsection{Ageing invariance}

For applications to ageing, we must consider to so-called {\em ageing algebra} 
\BEQ
\mathfrak{age}_1 :=\langle X_{0,1},Y_{-\frac{1}{2},\frac{1}{2}},M_0\rangle
\subset \mathfrak{sch}_1
\EEQ 
(without time-translations) 
which is a true subalgebra of $\mathfrak{sch}_1$. The generators $Y_m, M_n$
retain their form, but the generators $X_n$ only exist for $n\geq 0$. 
They now read \cite{Pico04,Henk06a}
\BEQ \label{gl:Xnext}
\hspace{-2truecm}
X_n = -t^{n+1}\partial_t - \frac{n+1}{2} t^n r\partial_r
-\frac{(n+1)n}{4}{\cal M}t^{n-1} r^2 - \frac{x}{2}(n+1)t^n -\xi n t^n
\;\; ; \;\; n\geq 0
\EEQ
where $\xi$ is a new quantum number associated with the field $\phi$ on 
which the generators $X_n$ act. When $\xi\ne 0$, the generators $X_n$ are only 
part of a representation of $\mathfrak{age}_1$ but not of $\mathfrak{sch}_1$. 
This is only possible for systems out of a stationary state (otherwise, 
the requirement of time-translation invariance
and $[X_1,X_{-1}]=2X_0$ would lead to $\xi=0$). With respect to (\ref{2:Sinv}),
only one commutator changes and we have
$[{\cal S},X_1] = -2t{\cal S} -(2x+2\xi-1)M_0$, resulting in the weaker
condition $x+\xi=1/2$ in order that the representation of $\mathfrak{age}_1$ 
can act as a dynamical symmetry. It can be shown that in 
the scaling form eqs.~(\ref{gl:sR},\ref{1:R}) 
of the two-time autoresponse function the difference $a'-a$ is related to
$\xi$ (and  $\wit{\xi}$ of the response field $\wit{\phi}$) and vanishes
if $\xi+\wit{\xi}=0$. 

The meaning of this new quantum number $\xi$ of a quasiprimary scaling 
operator becomes apparent when we consider the 
extension \cite{Henk94,Henk06a} of $\mathfrak{age}_1$ to the 
infinite-dimensional algebra $\mathfrak{av}$ spanned
by $(X_n)_{n\in\mathbb{N}_0}$, $(Y_m)_{m\in\mathbb{Z}+1/2}$ and 
$(M_n)_{n\in\mathbb{Z}}$, such that eq.~(\ref{gl:5:2:SchAlg}) remains valid. 
It is well-known that $Y_m$ and $M_n$ generate
time-dependent translations (which an additional phase) and phase shifts,
respectively. On the other hand, the generators $(X_n)_{n\geq 0}$ 
eq.~(\ref{gl:Xnext}) are the 
infinitesimal generators of the transformation 
$(t,\vec{r})\mapsto (t',\vec{r}')$ where
\BEQ 
t=\beta(t') \;\; , \;\; \vec{r} = \vec{r}' \sqrt{\frac{\D\beta(t')}{\D t'}\,}
\EEQ
such that $\beta(0)=0$ and the scaling operator $\phi$ transforms 
as \cite{Henk06a}
\BEQ \hspace{-1.7truecm}
\phi(t,\vec{r}) =\left(\frac{\D\beta(t')}{\D t}\right)^{-x/2} 
\left(\frac{\D \ln\beta(t')}{\D \ln t'}\right)^{-\xi} 
\exp\left[-\frac{{\cal M}{\vec{r}'\,}^2}{4}
\frac{\D}{\D t'}\left(\ln\frac{\D\beta(t')}{\D t'}\right)
\right]\,
\phi'(t',\vec{r}') \;
\EEQ
This transformation is different from the one of a 
Schr\"odinger quasiprimary operator
if $\xi\ne 0$ but it suggests to define the scaling operator
\BEQ \label{2:Pp}
\Phi(t,\vec{r}) := t^{-\xi}\,\phi(t,\vec{r})
\EEQ 
which is indeed quasiprimary
under the ageing algebra $\mathfrak{av}$ and even the analogous extension
$\mathfrak{sv}$ of the Schr\"odinger algebra $\mathfrak{sch}_1$, 
but with a modified  scaling dimension 
$x_{\Phi} = x_{\phi}+2\xi_{\phi}$ \cite{Henk06a}. 

Therefore, in order to describe the local scale-invariance of ageing systems
with $z=2$, it is enough to study first Schr\"odinger-quasiprimary operators
$\Phi$ and only at the end, one goes over to ageing scaling operators $\phi$
using eq.~(\ref{2:Pp}).

\subsection{Conformal invariance of the free diffusion equation}

Important information will come from {\it considering the `mass' $\cal M$ not
as some  constant, but rather as an additional variable}. It is useful
to work with the Fourier transform of the
field and of the generators with respect to $\cal M$ and to define a new
field $\psi$ as follows \cite{Giul96,Henk03}
\BEQ \label{2:gl:phipsi}
\phi(t,\vec{r}) = \phi_{\cal M}(t,\vec{r}) = \frac{1}{\sqrt{2\pi}} 
\int_{\mathbb{R}} \!\D\zeta\, e^{-\II {\cal M}\zeta}\, \psi(\zeta,t,\vec{r})
\EEQ
Provided $\lim_{\zeta\to\pm\infty}\psi(\zeta,t,\vec{r})=0$, 
the diffusion equation
becomes 
\BEQ \label{2:gl:Schr}
2\II \frac{\partial^2 \psi}{\partial t\partial\zeta} + 
\frac{\partial^2 \psi}{\partial \vec{r}^2} =0
\EEQ
which for brevity we shall also call a diffusion/Schr\"odinger equation. 
In $d=1$ the generators read 
(an eventual generalization to $\mathfrak{age}_1$ or $\mathfrak{av}$ 
is straightforward) 
\BEA
X_n &=& -t^{n+1}\partial_t - \frac{n+1}{2}\ t^n r\partial_r 
- (n+1)\frac{x}{2} t^n
+\II \frac{n(n+1)}{4} t^{n-1} r^2 \partial_{\zeta} \nonumber \\
Y_m &=& -t^{m+1/2} \partial_r + \II \left( m+\frac{1}{2}\right) t^{m-1/2}
 r \partial_{\zeta}
\nonumber \\
M_n &=&  \II t^n \partial_{\zeta}
\label{gl:Schr2}
\EEA
This change of variables trades the complicated phases acquired by the
field $\phi$ under a Schr\"odinger/ageing transformation 
for a time-dependent translation of the new internal coordinate
$\zeta$. For the Galilei-transformation $Y_{1/2}$ this was first observed by 
Giulini \cite{Giul96}.

The passage to the new variable $\zeta$ implies an extension of the dynamical 
symmetry algebra \cite{Henk03} which we illustrate for $d=1$. 
Rewrite the physical coordinates $\zeta,t,r$ as the components of a 
three-dimensional vector $\vec{\xi}=(\xi_{-1},\xi_0,\xi_1)$ where 
\BEQ
t=\half (-\xi_0+\II \xi_{-1}) \;\; , \;\;
\zeta=\half (\xi_0+\II \xi_{-1}) \;\; , \;\;
r=\sqrt{\frac{\II}{2}\,} \xi_1
\EEQ 
and $\psi(\zeta,t,r)=\Psi(\vec{\xi})$. Then the diffusion equation 
(\ref{2:gl:Schr}) becomes a three-dimensional massless Klein-Gordon
equation 
\BEQ \label{3:gl:KG}
\partial_{\mu} \partial^{\mu} \Psi(\vec{\xi}) = 0 
\EEQ
The Lie algebra of the maximal kinematic group of this equation is the 
conformal algebra $\mathfrak{conf}_{3}$, with generators
\BEA
P_{\mu} &=& \partial_{\mu} \nonumber \\
M_{\mu\nu} &=& \xi_{\mu}\partial_{\nu} - \xi_{\nu}\partial_{\mu}\nonumber\\
K_{\mu} &=& 2 \xi_{\mu}\xi^{\nu}\partial_{\nu} - 
\xi_{\nu}\xi^{\nu}\partial_{\mu} +2x \xi_{\mu} \label{3:gl:konfG} \\
D &=& \xi^{\nu}\partial_{\nu} +x \nonumber 
\EEA
($\mu,\nu\in\{-1,0,1\}$) 
which represent, respectively, translations, rotations, special transformations
and the dilatation. Hence the generators of ${\mathfrak{sch}}_1$ are linear 
combinations (with complex coefficients) of the above ${\mathfrak{conf}}_3$ 
generators. Hence, for any $d\geq 1$, 
{\it one has an inclusion of the complexified Lie algebra   
$({\mathfrak{sch}}_d)_{\C}$ into 
$({\mathfrak{conf}}_{d+2})_{\C}$} \cite{Burd73}. 
Explicitly, for $d=1$ 
\BEA
X_{-1}=\II (P_{-1}-\II P_0) \;\; , \;\;
X_0=-\half D+{\II\over 2} M_{-1 0}\;\; , \;\; 
X_1=-{\II\over 4} (K_{-1}+\II K_0) 
\nonumber\\
Y_{-1/2}=-\sqrt{\frac{2}{\II}\,} P_1 \;\; , \;\; 
Y_{\half}=-\sqrt{\frac{\II}{2}\,} (M_{-1 1}+\II M_{0 1})\;\; , \;\; 
M_0=P_{-1}+\II P_0.
\EEA
The four remaining generators needed to 
get the full conformal Lie algebra $({\mathfrak{conf}}_3)_{\C}$
can be taken in the form 
\BEA
N_{\;\,} =-t\partial_t +\zeta \partial_{\zeta} & &  
\mbox{\rm ~time-phase symmetry} 
\nonumber \\
V_- = 
-\zeta \partial_r+\II r\partial_t & &  
\mbox{\rm ~`dual' Galilei transformation} 
\nonumber \\
W_{\;\,} = 
-\zeta^2 \partial_{\zeta}-\zeta r\partial_r+{\II\over 2} r^2 \partial_t
-x\zeta \nonumber & & 
\mbox{\rm ~`dual' special transformation} \\
V_+ = 
-2tr\partial_t-2\zeta r\partial_{\zeta}-(r^2+2\II\zeta t)\partial_r-2xr 
& &  
\mbox{\rm ~transversal inversion}.
\EEA
The generators $V_-$ and $W$ are, up to constant coefficients, the 
complex conjugates of $Y_{1/2}$ and $X_1$, respectively, in the 
coordinates $\xi^{\mu}$, hence their names. The complex conjugation becomes
the exchange $t\leftrightarrow\zeta$ in the physical coordinates $(\zeta,t,r)$.

\begin{figure}
\epsfxsize=80mm
\centerline{\epsffile{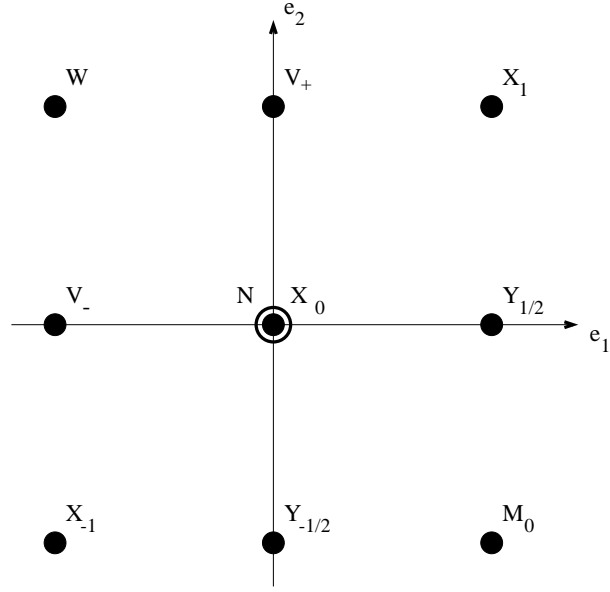}}
\caption{Roots of the complex simple Lie algebra $B_2$ and their relation 
with the generators of the
Schr\"odinger algebra $\mathfrak{sch}_1$ and the conformal algebra 
$\mathfrak{conf}_3$. The double circle in the centre
denotes the Cartan subalgebra $\mathfrak{h}$. 
\label{Abb1}}
\end{figure}

These results are illustrated in the root diagramme figure~\ref{Abb1} 
\cite{Henk03}. It can be shown \cite{Knap86} 
that the generators considered above are roots of the complex
Lie algebra $B_2 \cong (\mathfrak{conf}_3)_{\mathbb{C}}$. Then to each 
root ${\cal X}\in B_2$ one may associate a two-dimensional vector 
$\mathfrak{x}\in \mathbb{Z}^2$ and linking the origin to one of the points in 
the root diagramme such that forming the commutator 
$[{\cal X}_1, {\cal X}_2] = c_{12,3} {\cal X}_3$ corresponds 
to vector addition $\mathfrak{x}_1+\mathfrak{x}_2=\mathfrak{x}_3$ and 
$c_{12,3}=0$ if the result of that addition falls outside the root diagramme. 

Since possible isomorphisms are described by the Weyl group which is 
generated by the simple symmetries in figure~\ref{Abb1} 
$w_1: (e_1,e_2)\mapsto (-e_2,-e_1)$ and
$w_2: (e_1,e_2)\mapsto (-e_1,e_2)$ one may readily write down the list of 
non-isomorphic maximal subalgebras of $B_2$. These are \cite{Henk03}
\begin{enumerate}
\item
the  algebra 
$\wit{\mathfrak{age}}_1 := \mathfrak{age}_1 \oplus \mathbb{C} N$, obtained
from $\mathfrak{age}_1$ by adding the generator $N$.
This algebra is known as the
{\it minimal standard parabolic subalgebra}.\footnote{A parabolic
subalgebra is defined as the sum
of the Cartan subalgebra and of the positive root spaces \cite{Knap86}.}
\item
the extension 
$\wit{\mathfrak{sch}}_1 :={\mathfrak{sch}}_1\oplus \mathbb{C} N$ of the 
Schr\"odinger algebra which is also a  
parabolic subalgebra.
\item
the parabolic subalgebra 
algebra $\wit{\mathfrak{alt}}_1 := \mathfrak{alt}_1 \oplus \mathbb{C} N$
where $\mathfrak{alt}_1 := \langle D,Y_{-1/2}, M_0,Y_{1/2},V_+,X_1\rangle$.
See \cite{Henk06h} for a study of its representation theory.
\end{enumerate}
Against often-read statements, which go back to \cite{Baru73}, 
$\mathfrak{sch}_1$ cannot be obtained through
group contraction from taking a non-relativistic limit of the conformal 
algebra $\mathfrak{conf}_3$. Rather, this limit yields a projection
$\mathfrak{conf}_3 \to \mathfrak{alt}_1$ \cite{Henk03}. On the other hand,
$\mathfrak{alt}_1$ and its infinite-dimensional extension can be obtained
from a group contraction, realized as  a non-relativistic limit, 
of a pair of commuting Virasoro algebras \cite{Henk06h}. 

These new generators are indeed dynamical symmetries of the free 
diffusion/Schr\"odinger equation. To see this, consider the
$1D$ Schr\"odinger operator
\BEQ
{\cal S} = M_0 X_{-1} - Y_{-1/2}^2 = 2\II \partial_{\zeta}\partial_t 
+ \partial_r^2
\EEQ
It is straightforward to check that 
\BEA
\left[ {\cal S}, V_{-} \right] &=& 0 \:=\: \left[ {\cal S}, N \right]
\nonumber \\
\left[ {\cal S}, V_{+} \right] &=& 2(1-2x)\partial_t -4r{\cal S} 
\nonumber \\ 
\left[ {\cal S}, W \right] &=& \II(1 - 2x)\partial_r - 2\zeta {\cal S}
\label{3:gl:Kommu}
\EEA
which complement eq.~(\ref{2:Sinv}). Hence, {\it any solution $\psi$ of the 
$1D$ Schr\"odinger equation ${\cal S}\psi=0$ 
with a scaling dimensions $x=1/2$ is mapped onto another 
solution of (\ref{2:gl:Schr})
under the action of the conformal algebra $(\mathfrak{conf}_3)_{\C}$}. 
Of course, the space-time realization of this conformal symmetry is 
unconventional in the sense that space-time rotations with $\cal M$ fixed 
do {\em not} occur. 

A similar extension can be made for higher spin fields. The simplest example
treats the relationship of the non-relativistic L\'evy-Leblond equations to the
massless Dirac equation \cite{Henk06g}. 

For extensions to $d>1$ and the emerging mathematical theory of central
extensions of the Schr\"odinger-Virasoro algebra $\mathfrak{sv}$ and its 
vertex operators we refer to the litterature \cite{Roge06,Unte07}. 

Given the {\em conformal} invariance of the free Schr\"odinger equation, 
the form
of the two- and three-point functions is well-known \cite{Poly70}. Transforming
these results back to a fixed value of $\cal M$, one recovers the two-time
and three-time response functions, together with an explicit proof of 
their physically required causality properties \cite{Henk03}. 

\subsection{Reduction from Langevin to deterministic equations}

So far, we have considered the dynamical symmetries of simple diffusion
equations but we should really have been interested in the possible symmetries
of stochastic Langevin equations. For a non-conserved order-parameter (model A)
one usually writes \cite{Hohe77} 
\BEQ \label{gl:5:Langevin}
2{\cal M} \frac{\partial\phi}{\partial t} = 
\Delta \phi - \frac{\delta {\cal V}[\phi]}{\delta \phi} +\eta
\EEQ
where $\cal V$ is the Ginzburg-Landau potential and $\eta$
is a gaussian noise which describes the coupling to an external heat-bath and
the initial distribution of $\phi$. At first sight, there appear to be no
non-trivial symmetries (besides the obvious translation and rotation
invariance  and possibly dynamical scaling), since the noise $\eta$ does
break Galilei-invariance. This can be seen from an easy calculation, but in
order to understand this physically, consider a magnet which
is a rest with respect to a homogeneous heat-bath at temperature $T$. 
If the magnet is moved with a constant velocity with respect to 
the heat-bath, the effective
temperature will now appear to be direction-dependent, and the heat-bath
is no longer homogeneous. 

Still, there  are hidden non-trivial symmetries which come about as follows
\cite{Pico04}: {\em split the Langevin equation into a
`deterministic' part with non-trivial dynamic symmetries and a `noise' part 
and then show using these symmetries that all averages can be reduced exactly 
to averages within the deterministic, noiseless theory}. Technically, one
first constructs in the standard fashion (Janssen-de Dominicis procedure)
\cite{deDo78,Jans92} the associated stochastic field-theory with action 
${\cal J}[\phi,\wit{\phi}]$ where $\wit{\phi}$ is the response field 
associated to 
the order-parameter $\phi$. Second, decompose the action into two parts
\BEQ \label{gl:5:JanDom}
{\cal J}[\phi,\wit{\phi}] = {\cal J}_0[\phi,\wit{\phi}] + {\cal J}_b[\wit{\phi}]
\EEQ
where  
\BEQ \label{gl:5:JanDomdet}
{\cal J}_0[\phi,\wit{\phi}]=
\int_{\mathbb{R}_+\times\mathbb{R}^d}\!\D t\D\vec{r}\; 
\wit{\phi}\left(2{\cal M}\partial_t\phi-\Delta\phi+
\frac{\delta{\cal V}}{\delta\phi}\right)
\EEQ 
contains the terms coming from the `deterministic' part of
the Langevin equation ($\cal V$ is the self-interacting `potential') whereas
\BEQ \label{gl:5:JanDombruit}
\hspace{-1truecm}
{\cal J}_b[\wit{\phi}] = 
-T \int_{\mathbb{R}_+\times\mathbb{R}^d}\!\D t\D\vec{r}\: 
\wit{\phi}(t,\vec{r})^2 
-\frac{1}{2}
\int_{\mathbb{R}^{2d}} \!\D\vec{r}\,\D\vec{r}'\:  
{\wit{\phi}}(0,\vec{r})a(\vec{r}-\vec{r}'){\wit{\phi}}(0,\vec{r}')
\EEQ
contains the `noise'-terms coming from (\ref{gl:5:Langevin}) \cite{Jans92}. 
It was assumed here that $\langle \phi(0,\vec{r})\rangle =0$ and
$a(\vec{r})$ denotes the initial two-point correlator 
\BEQ \label{2:aini}
a(\vec{r}) := C(0,0;\vec{r}+\vec{r}',\vec{r}') = 
\langle \phi(0,\vec{r}+\vec{r}') \phi(0,\vec{r}')\rangle =a(-\vec{r})
\EEQ
where the last relation follows from spatial translation/rotation-invariance 
which we shall admit throughout.

It is instructive to consider briefly the case of a free field, where
${\cal V}=0$. Variation of (\ref{gl:5:JanDom}) with respect to $\wit{\phi}$ 
and $\phi$, respectively, then leads to the equations of motion
\BEQ \label{gl:5:equamovi}
2{\cal M}\partial_t \phi = \Delta \phi + T \wit{\phi} \;\; , \;\;
-2{\cal M}\partial_t\wit{\phi} = \Delta\wit{\phi}
\EEQ
The first one of those might be viewed as a Langevin equation if $\wit{\phi}$
is interpreted as a noise. Comparison of the two equations of motion 
(\ref{gl:5:equamovi}) shows that if the order-parameter $\phi$ is 
characterized by
the `mass'\index{mass} $\cal M$ (which by physical convention is positive), 
then the associated response field $\wit{\phi}$ is 
characterized by the {\em negative} mass $-{\cal M}$. 
This characterization remains valid beyond free fields. 

We now concentrate on actions ${\cal J}_0[\phi,\wit{\phi}]$ which
are Galilei-invariant. This means that if $\langle .\rangle_0$ denotes the
averages calculated only with the action ${\cal J}_0$, 
the Bargman superselection rules \cite{Barg54}
\BEQ \label{gl:5:Bargman}
\left\langle \underbrace{\phi\ldots\phi}_n \; \underbrace{\wit{\phi}\ldots
\wit{\phi}}_m \right\rangle_0 \sim \delta_{n,m}
\EEQ
hold true. It follows that both response and correlation functions can be
exactly expressed in terms of averages with respect to the deterministic part
alone. The simplest example is the two-time response function (we suppress
for notational simplicity the spatial coordinates) \cite{Pico04}
\BEQ
\hspace{-1truecm} 
R(t,s) = \left.\frac{\delta\langle \phi(t)\rangle}{\delta h(s)}\right|_{h=0}
= \left\langle \phi(t) \wit{\phi}(s) \right\rangle
= \left\langle \phi(t) \wit{\phi}(s)\, e^{-{\cal J}_b[\wit{\phi}]}\right\rangle_0
= \left\langle \phi(t) \wit{\phi}(s) \right\rangle_0 
\label{gl:5:Rrauschlos}
\EEQ
where the `noise' part of the action was included in the
observable, the exponential expanded and the Bargman superselection rule
(\ref{gl:5:Bargman}) was used. In other words, 
{\em the two-time response function
does not depend explicitly on the `noise' at all} and this explains why
the prediction (\ref{1:R}) which was derived from the dynamical symmetries of
a deterministic equation successfully describes the response functions 
calculated in stochastic models, as reviewed recently in \cite{Henk07a,Henk07b}.

The correlation function is reduced similarly \cite{Pico04}
\BEA
\lefteqn{ 
C(t,s;\vec{r}) = 
T \int_{\mathbb{R}_+\times\mathbb{R}^d}\!\D u\D\vec{R}\: 
\left\langle \phi(t,\vec{r}+\vec{y})\phi(s,\vec{y})\wit{\phi}(u,\vec{R})^2
\right\rangle_0 
}
\nonumber \\
& & +\frac{1}{2} \int_{\mathbb{R}^{2d}}\!\D\vec{R}\D\vec{R}'\: 
a(\vec{R}-\vec{R}')
\left\langle \phi(t,\vec{r}+\vec{y})\phi(s,\vec{y})
\wit{\phi}(0,\vec{R})\wit{\phi}(0,\vec{R}') \right\rangle_0
\label{gl:5:Crauschlos} 
\EEA
Only terms which depend
explicitly on the `noise' remain -- recall the vanishing of 
the `noiseless' two-point function 
$\langle \phi(t)\phi(s)\rangle_0=0$ because of the Bargman superselection rule. 
Eq.~(\ref{gl:5:Crauschlos}) is the basis for our later explicit calculation
of $C(t,s)$. 

\subsection{Extensions}

For the later applications to phase-ordering, we need the following extension
of (\ref{gl:5:Langevin})
\BEQ \label{2:LangevinE}
2{\cal M} \frac{\partial\phi}{\partial t} = 
\Delta \phi  -\frac{\delta {\cal V}[\phi]}{\delta \phi} - v(t) \phi +\eta
\EEQ
where the time-dependent potential $v(t)$ arises naturally in spherical models.
In more general models, it might be viewed as a Lagrange multiplier to enforce
that $C(t,t)$ is finite. Through the gauge transformation 
\BEQ \label{2:jauge}
\phi(t,\vec{r}) \mapsto \phi(t,\vec{r})\, k(t) \;\; ; \;\;
k(t) := \exp\left(-\frac{1}{2{\cal M}} \int_0^t \!\D t'\: v(t') \right) 
\EEQ
eq.~(\ref{2:LangevinE}) is reduced to (\ref{gl:5:Langevin}) \cite{Pico04}. 

Finally, we consider semi-linear equations (\ref{gl:5:Langevin}) where
${\cal V}={\cal V}[\phi,\wit{\phi}]$. Here the essential ingredient is
to recognise that the coupling constant $g$ of a non-linear term will in general
have some scaling dimension $y_g$. This implies in turn that under local
scale-transformations $g$ must be transformed as well and consequently, the
infinitesimal generators have to be generalized accordingly. Standard methods
\cite{Boye76} then allow to compute the Schr\"odinger-invariant 
semi-linear equations and one has:
\begin{enumerate}
\item for fixed masses, the invariant semi-linear Schr\"odinger equations
are of the form
\BEQ
{\cal S}\phi = \phi \left( \phi \wit{\phi}\right)^{1/x} 
f\left( g^x \left( \phi \wit{\phi}\right)^{y_g} \right)
\EEQ
where $x$ is the scaling dimension of $\phi$ and $f$ an arbitrary
differentiable function \cite{Baum05b}. 
\item for variable masses, the $1D$ invariant semi-linear Schr\"odinger
equations with real-valued solutions are of the form
\BEQ
{\cal S} \psi = \psi^5 \bar{f}\left( g \psi^{4 y_g}\right)
\EEQ
where $\bar{f}$ is an arbitrary differentiable function \cite{Stoi05}. 
We point out that in this way one can obtain Schr\"odinger-invariant diffusion
equations with real-valued solutions, at the expense that after inverse Laplace
transformation with respect to $\zeta$ products $\psi \cdot \psi \cdot \ldots$
are replaced by convolutions $\phi * \phi * \ldots$ with respect to $\cal M$. 
\end{enumerate}
Slightly more general forms are possible if only ageing-invariance is
required \cite{Baum05b,Stoi05}. 

\section{Calculation of two-time autocorrelation functions}

\subsection{Schr\"odinger-invariant autocorrelators} 

We now turn to the explicit calculation of the two-time autocorrelation
function, starting from the exact reduction formula eq.~(\ref{gl:5:Crauschlos}). 
As initial condition, we use a fully disordered initial state,
with $a(\vec{r}) = a_0 \delta(\vec{r})$, see eq.~(\ref{2:aini}). 
Then the calculation reduces to find a three-point response function
$\langle \phi(t) \phi(s)\wit{\phi}^2(u)\rangle_0$ within the deterministic
part of the theory and local scale-invariance will become useful here. 
At first sight, since there no time-translation-invariance, it might
appear that the only dynamical symmetry available would be $\mathfrak{age}_d$,
with rather limited predictive capacities. It is better to proceed as
follows \cite{Henk04b,Henk06a}: 
\begin{enumerate}
\item we break time-translation-invariance {\em explicitly} by considering a
Langevin equation of the form (\ref{2:LangevinE}) with a time-dependent
potential $v(t)$. We can reduce this to the standard form (\ref{gl:5:Langevin}) 
via the gauge transformation (\ref{2:jauge}) 
\BEQ \label{3:kvf}
k(t) \sim t^{\digamma} \mbox{\rm ~~ or equivalently ~~} 
v(t) = -\frac{2{\cal M}\digamma}{t}
\EEQ
\item recall that for ageing-invariance 
the scaling operators $\phi$ (with scaling dimension $x$) 
do not transform as conventional
quasi-primary scaling operators, but are related via (\ref{2:Pp}) to
{\it bona fide} quasiprimary scaling operators $\Phi$, with scaling
dimension $x+2\xi$. 
\item we then use full Schr\"odinger-invariance for the calculation of
the required three-point function 
$\langle\Phi(t)\Phi(s)\wit{\Phi}^2(u)\rangle_0$.
\item since this procedure still leaves a scaling function of a single
variable undetermined, we use the further extension from Schr\"odinger
to $(d+2)$-dimensional conformal invariance. This requires to consider 
the masses as further variables, however. 
\end{enumerate} 

For illustration, we briefly consider the autoresponse function
$R(t,s)=\langle \phi(t)\wit{\phi}(s)\rangle_0$. According to the procedure 
outlined above, $\phi$ and $\wit{\phi}$ are characterized by the exponents
$x,\xi$ and $\wit{x},\wit{\xi}$, respectively, in addition to the parameter
$\digamma$ coming from the gauge transformation (\ref{2:jauge},\ref{3:kvf}). 
This leads to \cite{Henk06a} 
\BEA
R(t,s) &=& \left\langle t^{\xi} \Phi(t,\vec{r}) \,
s^{\wit{\xi}}\, \wit{\Phi}(s,\vec{r}) \right\rangle_0 
\nonumber \\
&=& t^{\xi} s^{\wit{\xi}} \, \frac{k(t)}{k(s)}\, 
(t-s)^{-(x+2\xi)} \Theta(t-s)\,
\delta_{x+2\xi,\wit{x}+2\wit{\xi}} 
\nonumber \\
&=& s^{-(x+\wit{x})/2} \left(\frac{t}{s}\right)^{\xi+\digamma} 
\left(\frac{t}{s}-1\right)^{-x-2\xi} \Theta(t-s)\, 
\delta_{x+2\xi,\wit{x}+2\wit{\xi}}
\EEA
which reproduces (\ref{1:R}), including its causality, 
with the following identification of exponents (we use $z=2$ here)
\BEQ \label{3:Rexpos}
\lambda_R = 2(x+\xi-\digamma) \;\; , \;\;
a'-a = \xi +\wit{\xi} \;\; , \;\;
1+a = \half (x+\wit{x})
\EEQ
Alternatively, the gauge exponent $\digamma$ may be expressed as
$\digamma=\half (\wit{x}-x) +\wit{\xi}-\xi$. 

Similarly, the autocorrelation function can be written as, where the
composite field $\wit{\phi}^2$ is characterized by $2\wit{x}_2$ and 
$2\wit{\xi}_2$
\newpage \typeout{*** ici un saut de page !! ***}
\BEA
\hspace{-2.0truecm}C(t,s) &\hspace{-1.0truecm}=& 
\frac{a_0}{2}\int_{\mathbb{R}^d} 
\!\D\vec{R}\: R_{0}^{(3)}(t,s,\tau_{\rm ini};\vec{R})
+\frac{T}{2{\cal M}}
\int_0^{\infty} \!\D u\int_{\mathbb{R}^d}\!\D\vec{R}\: 
R_{0}^{(3)}(t,s,u;\vec{R})
\label{glC1}
\\
\hspace{-2.0truecm}R_{0}^{(3)}(t,s,u;\vec{r}) &\hspace{-0.0truecm}:=&
\left\langle\phi(t;\vec{y})
\phi(s;\vec{y}){\wit \phi}(u;\vec{r}+\vec{y})^2\right\rangle_0
\nonumber \\
&\:=& (ts)^{\xi} u^{2\wit{\xi}_2} \left\langle\Phi(t;\vec{y})
\Phi(s;\vec{y}){\wit \Phi}(u;\vec{r}+\vec{y})^2\right\rangle_0
\nonumber \\
&\:=& (ts)^{\xi} u^{2\wit{\xi}_2}\,\frac{k(t)k(s)}{k^{2}(u)} 
{\cal R}_{0}^{(3)}(t,s,u;\vec{r})
\label{glC2}
\EEA
Here $\tau_{\rm ini}$ is some small `initial' time which sets the 
beginning of the scaling regime. At a later stage, the limit 
$\tau_{\rm ini}\to 0$ should be taken. 

Throughout, we consider $\wit{\Phi}^2$ as a composite field with
scaling dimension $2\wit{x}_2 + 4\wit{\xi}_2$ (only for free fields, 
one would have $\wit{x}=\wit{x}_2$ and $\wit{\xi}=\wit{\xi}_2$). 
Schr\"odinger-invariance gives the three-point response function
${\cal R}_0^{(3)}$ for $v(t)=0$ \cite{Henk94}
\BEA
\hspace{-2.3truecm}{\cal R}_{0}^{(3)}(t,s,u;\vec{r}) &=& 
{\cal R}_{0}^{(3)}(t,s,u)
\exp\left[-\frac{{\cal M}}{2}\frac{t+s-2u}{(s-u)(t-u)}{\vec{r}}^{2} \right]
\Psi\left(\frac{t-s}{2(t-u)(s-u)}{\vec{r}}^{2}\right)
\nonumber \\
\hspace{-2.3truecm}{\cal R}_{0}^{(3)}(t,s,u) &=& \Theta(t-u)\Theta(s-u)\,
\left(t-u\right)^{-{\wit x}_2-2\wit{\xi}_2}
\left(s-u\right)^{-{\wit x}_2-2\wit{\xi}_2}
\left(t-s\right)^{-x-2\xi+{\wit x}_2+2\wit{\xi}_2}~
\label{glC3}
\EEA
where $\Psi=\Psi(\rho)$ is an arbitrary scaling function, to be determined. 
Of course, and we shall come back to this below, this is only applicable 
within the ageing regime as defined in eq.~(\ref{1:validite}). 
It is useful to write down the autocorrelation function in the form
$C(t,s)=C_{\rm th}(t,s)+C_{\rm pr}(t,s)$. Here the thermal part 
$C_{\rm th}$ and the preparation part $C_{\rm pr}$ (where the limit
$\tau_{\rm ini}\to 0$ was taken) are given by
\BEA
\hspace{-2.2truecm}C_{\rm th}(t,s) &=& C_{0,{\rm th}}\, s^{d/2+1-x-{\wit x}_2}
\:\left(\frac{t}{s}\right)^{\digamma}
\left(\frac{t}{s}-1\right)^{{\wit x}_2+2\wit{\xi}_2-x-2\xi-d/2}
\nonumber \\
\hspace{-2.2truecm}& & \times 
\int_{0}^{1}\!\,\D v\: v^{2\wit{\xi}_2-2\digamma}
\left[\left(\frac{t}{s}-v\right)
\left(1-v\right)\right]^{d/2-{\wit x}_2-2\wit{\xi}_2}
\vec{\Psi}\left(\frac{t/s+1-2v}{t/s-1}\right)
\nonumber \\
\hspace{-2.2truecm}C_{\rm pr}(t,s) &=& C_{0,{\rm pr}}\,
s^{d/2+2\digamma-{\wit x}_2-2\wit{\xi}_2-x}
\nonumber \\
\hspace{-2.2truecm}& & \times
\left(\frac{t}{s}\right)^{d/2+\xi+\digamma-\wit{x}_2-2\wit{\xi}_2}
\left(\frac{t}{s}-1\right)^{{\wit x}_2-x+2\wit{\xi}_2-2\xi-d/2}
\vec{\Psi}\left(\frac{t/s+1}{t/s-1}\right)
\nonumber \\
\hspace{-2.2truecm}\vec{\Psi}(w) &:=& \int_{\mathbb{R}^d}
\!\D\vec{R}\:\exp\left[-\frac{{\cal M}w}{2}\,{\vec R}^2\right]
\Psi\left({\vec R}^{2}\right)
\label{eqC6}
\EEA   
and we have explicitly used $s<t$. Here $C_{0,{\rm th}}$ is a normalization
constant proportional to the temperature $T$ and $C_{0,{\rm pr}}$ is a
rescaled normalization depending on the initial time-scale $\tau_{\rm ini}$.

For the interpretation of these forms, we concentrate on simple magnets
relaxing towards  equilibrium stationary states. 
{}From renormalization arguments \cite{Bray94,Cala05} one expects that
the thermal part $C_{\rm th}$ will be leading one for quenches {\em onto}
the critical temperature $T=T_c$ while the preparation part $C_{\rm pr}$
should be dominant one for quenches into the ordered phase  ($T<T_c$). 
Comparison with the expected scaling forms (\ref{gl:sC},\ref{1:CR}) then give
\begin{enumerate}
\item \underline{\bf $T=T_c$:} 
for relaxation to equilibrium, one has\footnote{If one
attempts to apply this kind or argument to systems relaxing to critical 
non-equilibrium steady states, one must take into account that $a\ne b$ 
in general, see \cite{Henk07a} and references therein.} $a=b$ and from 
(\ref{eqC6}) we read off
\BEQ
a = b = x+\wit{x}_2 -1 -\frac{d}{2} \;\; , \;\; 
\lambda_C = 2 (x+\xi-\digamma)
\EEQ
The autocorrelation function scaling function becomes, up to normalization
\BEA
\hspace{-2.2truecm}f_C^{({\rm th})}(y) &=& y^{1+a'-\lambda_C/2}
\left(y-1\right)^{b-2a'+2\mu-1} 
\nonumber \\
\hspace{-2.2truecm}& & 
\times \int_0^1 \!\D v\: v^{\lambda_C+2\mu-2a'-2} \left[ \left(y-v\right)
\left(1-v\right)\right]^{a'-b-2\mu} 
\vec{\Psi}\left(\frac{y+1-2v}{y-1}\right)
\label{3:fCcrit}
\EEA
where $\mu:=\xi+\wit{\xi}_2$ is a further parameter whose value is not 
predicted by the theory. 
\item \underline{\bf $T<T_c$:} 
one has $b=0$ and from the asymptotic scaling function we find 
\BEQ
\lambda_C = 2 (x+\xi-\digamma)
\EEQ
The scaling function reads simply, up to normalization
\BEQ \label{3:fCpo}
f_C^{({\rm pr})}(y) = y^{\lambda_C/2} (y-1)^{-\lambda_C} 
\vec{\Psi}\left( \frac{y+1}{y-1} \right)
\EEQ
Since $\wit{x}_2-x+2\wit{\xi}_2-2\xi=\half d-\lambda_C\leq 0$ because of the
Yeung-Rao-Desai inequality \cite{Yeun96}, this exponent combination will only
vanish for free field. This illustrates again that $\wit{x}_2$ and 
$\wit{\xi}_2$ must be treated as independent quantities and cannot {\it a 
priori} be related to other exponents. 
\end{enumerate}

Comparing the above results for $\lambda_C$ with eq.~(\ref{3:Rexpos}) 
giving $\lambda_R$,
we see that {\em for short-ranged initial conditions and a deterministic part
${\cal J}_0$ of the action which is Schr\"odinger-invariant, 
the autocorrelation and autoresponse exponents are equal, 
$\lambda_C=\lambda_R$} \cite{Pico04}. This well-known result \cite{Bray94} 
can therefore also be understood as a consequence of a dynamical symmetry.

Schr\"odinger-invariance by itself does not determine the form of 
$\vec{\Psi}(\rho)$. However, the related three-point response function 
should be non-singular as $t-s\to 0$. This requirement fixes 
the asymptotic behaviour for $w\to\infty$ \cite{Pico04,Henk06a}
\BEQ \label{8:gl:2}
\vec{\Psi}(w) \sim  w^{-\vph} \;\; , \;\; \vph = \left\{
\begin{array}{ll} \lambda_C       & \mbox{\rm ~~;~ if $T<T_c$ and $b=0$} \\
                  1+2a'-b-2\mu    & \mbox{\rm ~~;~ if $T=T_c$ and $a=b$} 
\end{array} \right.
\EEQ
This suggests the following approximate forms of the scaling function, 
with $y=t/s$
\BEQ \label{8:gl:3}
\renewcommand{\arraystretch}{1.3}
\hspace{-1.8truecm}f_C(y) \approx \left\{ \begin{array}{ll}
\left[ (y+1)^2/(4y)\right]^{-\lambda_C/2} 
  & \mbox{\rm ~~;~ if $T<T_c$} \\
y^{1+a'-\lambda_C/2} \int_0^1 \!\D v\, 
v^{\lambda_C-2-2a'-2\mu} \left[(y-v)(1-v)\right]^{a'-b-2\mu} 
  & \mbox{\rm ~~;~ if $T=T_c$} \\
  \hspace{1.8truecm}\times (y+1-2v)^{b-2a'-1+2\mu} & 
\end{array} \right.
\renewcommand{\arraystretch}{1.0}
\EEQ
At least, these forms are consistent with the required asymptotic behaviour of
$f_C(y)$ as $y\to\infty$. For free-field theories (\ref{8:gl:2}) 
holds for all values of $w$ and then (\ref{8:gl:3}) becomes exact. 

\subsection{Conformal invariance}

In order to find the last remaining scaling function $\vec{\Psi}$, we now
propose to extend  Schr\"odinger-invariance further towards conformal
invariance in $d+2$ dimensions as described in section~2. This implies that
we must consider the `mass' ${\cal M}$ of the scaling operator $\Phi$ as
a further variable. Consider three Schr\"odinger-quasiprimary scaling
operators 
$\Phi_{\alpha}=\Phi_{\alpha}(t_{\alpha},\vec{r}_{\alpha},{\cal M}_{\alpha})$
where $\alpha=a,b,c$. Using the variables
\BEQ
\tau = t_a - t_c \;\; , \;\;
\sigma = t_b - t_c \;\; , \;\;
\vec{r} = \vec{r}_a -\vec{r}_c \;\; , \;\;
\vec{s} = \vec{r}_b -\vec{r}_c
\EEQ
the three-point function is
\BEQ
\langle \Phi_a\Phi_b\Phi_c\rangle_0 = \delta\left(\sum_{\alpha=a,b,c}
{\cal M}_{\alpha}\right) F(\tau,\sigma,\vec{r},\vec{s};
{\cal M}_a,{\cal M}_b,{\cal M}_c) 
\EEQ
where the delta function expresses the Bargman superselection rule 
and because of causality, $\tau>0, \sigma>0$. Schr\"odinger-invariance
almost fixes the form of $F$: 
\BEA
F &=& \tau^{-\gamma_1}\sigma^{-\gamma_2}(\tau-\sigma)^{-\gamma_3}
\exp\left[-\frac{{\cal M}_a}{2}\frac{\vec{r}^2}{\tau}
-\frac{{\cal M}_b}{2}\frac{\vec{s}^2}{\sigma}\right]
\Psi(\rho,{\cal M}_a,{\cal M}_b)
\\
\rho &=& \frac{(\vec{r}\sigma - \vec{s}\tau)^2}{2\tau\sigma(\tau-\sigma)}
\EEA
and
\BEQ
\gamma_1 = \half x_{ac,b} +\xi_{ac,b} \;\; ,\;\;
\gamma_2 = \half x_{bc,a} +\xi_{bc,a} \;\; ,\;\;
\gamma_3 = \half x_{ab,c} +\xi_{ab,c}
\EEQ
with $x_{ab,c}:=x_a+x_b-x_c$ and $\xi_{ab,c}:=\xi_a+\xi_b-\xi_c$. 

The last function $\Psi=\Psi_{abc}$ will now be found by extending invariance 
under $\mathfrak{sch}_d$ to invariance under the conformal algebra 
$\mathfrak{conf}_{d+2}$. For the explicit calculation, it is enough to 
consider the case $d=1$. From the root diagramme figure~\ref{Abb1} it is
clear that if one has invariance under the generators $N$ and $V_-$, then
invariance under the other generators follows. The explicit single-particle
form of the generators $N$ and $V_-$ is, for $d=1$
\BEA
N &=& -t\partial_t -1 -{\cal M}\partial_{\cal M}
\nonumber \\
V_- &=& -\II \partial_{\cal M}\partial_r +\II r\partial_t
\EEA  
Then the covariance of the three-point function under $N$ and $V_-$ leads
to
\BEA
\left( \tau\partial_{\tau} +\sigma\partial_{\sigma} +2
+{\cal M}_a\partial_{{\cal M}_a}+{\cal M}_b\partial_{{\cal M}_b}
\right) F &=& 0
\nonumber \\
\left(r\partial_{\tau} + s\partial_{\sigma} -
\partial_{{\cal M}_a}\partial_r - \partial_{{\cal M}_a}\partial_r
\right) F &=& 0
\EEA
In turn, these imply the following conditions for the as yet unknown
function $\Psi$
\BEA
\bigl[(2-\gamma_1-\gamma_2-\gamma_3) -\rho\partial_{\rho} 
+{\cal M}_a\partial_{{\cal M}_a}+{\cal M}_b\partial_{{\cal M}_b}
\bigr] \Psi &=& 0 
\label{3:eqPsi1}
\\
\left[ 
\frac{r}{\tau}\left[ 1-\gamma_1 + {\cal M}_a\partial_{{\cal M}_a} 
+\partial_{{\cal M}_a}\partial_{\rho} \right]
+ \frac{s}{\sigma}\left[ 1-\gamma_2 + {\cal M}_b\partial_{{\cal M}_b} 
+\partial_{{\cal M}_b}\partial_{\rho} \right] \right.& & 
\nonumber \\
~+ \left. \frac{r-s}{\tau-\sigma}\left[ -\gamma_3 -\rho\partial_{\rho}
-\partial_{{\cal M}_a}\partial_{\rho}
-\partial_{{\cal M}_b}\partial_{\rho} \right]
\right] \Psi &=& 0
\label{8:eqPsi2}
\EEA
In eq.~(\ref{8:eqPsi2}), it can be seen that if two of the inner
square brackets vanish when applied  to $\Psi$, the vanishing  of the
third one follows. Consequently, we have the following linear system of three
equations for the function $\Psi$ which depends on three variables
\BEA
\left[(2-\gamma_1-\gamma_2-\gamma_3) -\rho\partial_{\rho} 
+{\cal M}_a\partial_{{\cal M}_a}+{\cal M}_b\partial_{{\cal M}_b}
\right] \Psi &=& 0 
\nonumber \\
\left[ 1-\gamma_1 + {\cal M}_a\partial_{{\cal M}_a} 
+\partial_{{\cal M}_a}\partial_{\rho} \right] \Psi &=& 0 
\\
\left[ 1-\gamma_2 + {\cal M}_b\partial_{{\cal M}_b} 
+\partial_{{\cal M}_b}\partial_{\rho} \right] \Psi &=& 0
\nonumber 
\EEA
The first of these leads to the scaling form
\BEQ
\Psi(\rho,{\cal M}_a,{\cal M}_b) = \rho^{2-\gamma_1-\gamma-2-\gamma_3}\,
K\left( \frac{{\cal M}_a+{\cal M}_b}{2}\rho, 
\frac{{\cal M}_a-{\cal M}_b}{2}\rho\right)
\EEQ
and the other two give the following system of equations for $K=K(\eta,\zeta)$
\BEA
\hspace{-1.5truecm}\left[ 
2-\gamma_1-\gamma_2+(3-\gamma_1-\gamma_2-\gamma_3)\partial_{\eta}
+\eta\partial_{\eta}+\zeta\partial_{\zeta}+\eta\partial_{\eta}^2
+\zeta\partial_{\eta}\partial_{\zeta}\right] K(\eta,\zeta) &=& 0
\nonumber \\
\hspace{-1.5truecm}\left[ 
\gamma_2-\gamma_1+(3-\gamma_1-\gamma_2-\gamma_3)\partial_{\zeta}
+\eta\partial_{\zeta}+\zeta\partial_{\eta}+\zeta\partial_{\zeta}^2
+\eta\partial_{\eta}\partial_{\zeta}\right] K(\eta,\zeta) &=& 0
\EEA
which in principle can be solved through a double series in $\eta$ and $\zeta$.

For our purpose, namely to find the two-time autocorrelation function, we 
merely need
the special three-point function $\langle\Phi\Phi\wit{\Phi}^2\rangle_0$. 
In this case, we have
\BEQ
\gamma_1 = \gamma_2 = \wit{x}_2+2\wit{\xi}_2 \;\;,\;\;
\gamma_3 = x-\wit{x}_2+2\xi-2\wit{\xi}_2
\EEQ
${\cal M}_a={\cal M}_b={\cal M}$ and ${\cal M}_c = -2{\cal M}$. Hence
$\Psi=\Psi_{\Phi\Phi\wit{\Phi}^2}=\rho^{2-x-\wit{x}_2-2\xi-2\wit{\xi}_2}
K({\cal M}\rho,0)$ where, up to normalization, $K(\eta,0)$ is the solution
of
\BEQ
\left[ 2 -2\gamma_1 +(3-2\gamma_1-\gamma_3)\partial_{\eta}
+\eta\partial_{\eta}+\eta\partial_{\eta}^2 \right] K(\eta,0) = 0
\EEQ
This is readily solved in terms of confluent hypergeometric functions 
and the final result for the scaling function $\Psi$
reads, where $\psi_{0,1}$ are arbitrary constants
\BEA 
\hspace{-1.5truecm}\lefteqn{
\Psi(\rho,{\cal M},{\cal M}) = \psi_0\, 
\rho^{2-x-\wit{x}_2-2\xi-2\wit{\xi}_2}
{_1F_1}\left(2-2\wit{x}_2-4\wit{\xi}_2,3-x-\wit{x}_2-2\xi-2\wit{\xi}_2
;-{\cal M} \rho\right) 
}
\nonumber \\
\hspace{-1.5truecm}& & +\psi_1\, {\cal M}^{x+\wit{x}_2+2\xi+2\wit{\xi}_2-2}
{_1F_1}\left(x-\wit{x}_2+2\xi-2\wit{\xi}_2,x+\wit{x}_2+2\xi+2\wit{\xi}_2-1
;-{\cal M} \rho\right)
\label{3:PsiConf} 
\EEA

The main applications of eq.~(\ref{3:PsiConf}) 
will be to phase-ordering kinetics ($T<T_c$), 
where $z=2$ for simple magnets without disorder \cite{Bray94b} 
and furthermore $b=0$. Then 
\BEA 
\hspace{-1.5truecm}\lefteqn{
\Psi(\rho,{\cal M},{\cal M}) = \psi_0\, \rho^{\lambda_C-2a'-d/2}
{_1F_1}\left(2\lambda_C-d-2a',\lambda_C+1-2a'-d/2;-{\cal M} \rho\right) 
}
\nonumber \\
\hspace{-1.5truecm}& & +\psi_1\, {\cal M}^{2a'+d/2-\lambda_C}
{_1F_1}\left(\lambda_C-d/2,1+2a'+d/2-\lambda_C;-{\cal M} \rho\right)
\label{gl:3:Psipo} 
\EEA
The case $a=a'$ was already stated in \cite{Henk04b}. 
On the other hand, for non-equilibrium critical dynamics (quenches to $T=T_c$)
with dynamical exponent $z=2$ and also $a=b$ we have
\BEA 
\hspace{-1.5truecm}\lefteqn{
\Psi(\rho,{\cal M},{\cal M}) = \psi_0\, \rho^{1-b-2\mu-d/2}
{_1F_1}\left(2-d+2a'-2b-4\mu,2-b-2\mu-d/2;-{\cal M} \rho\right) 
}
\nonumber \\
\hspace{-1.5truecm}& & +\psi_1\, {\cal M}^{d/2+b+2\mu-1}
{_1F_1}\left(1-d/2+2a'-b-2\mu,d/2+b-2\mu;-{\cal M} \rho\right)
\label{gl:3:Psicr} 
\EEA
We point out that the previouly derived result eq.~(\ref{1:R}) for the response
functions does remain valid for this type of conformal invariance \cite{Henk03} 
and that the exponent $a'$ of the autoresponse function 
plays a r\^ole in fixing the form of $C(t,s)$.

Before we insert these explicit results into the scaling forms
eqs.~(\ref{3:fCpo}) and (\ref{3:fCcrit}), we should consider under which
physical conditions their derivation is valid. In particular, we used  from 
the outset dynamical scaling and and therefore restricted to the  
ageing regime eq.~(\ref{1:validite}). Especially the condition
$t-s\gg t_{\rm micro}$ must be satisfied, see \cite{Zipp00} for a careful 
discussion of this point. In our context, this means that for small
arguments $\rho\to 0$, the form of the function $\Psi(\rho)$ is
{\em not} given by local scale-invariance. Rather, for $\rho\ll 1$ one 
should expect that the two-time autocorrelation function $C(t,s)=C(s,t)$
should be symmetric in $t$ and $s$ and especially $C(t,s)$ should be
non-singular in the limit $t-s\to 0$. The requirement of the absence of
a singularity in $C(t,s)$ as $t-s\to 0$ is equivalent to an analogous 
requirement on the three-point function 
$\langle\Phi(t)\Phi(s)\wit{\Phi}^2(u)\rangle_0$ and leads to 
\BEQ
\Psi(\rho) \sim \rho^{x-\wit{x}_2+2\xi-2\wit{\xi}_2} \mbox{\rm ~~;~ as
$\rho\to 0$}
\EEQ
Specifically, for the two physical applications under  study, one recovers
eq.~(\ref{8:gl:2}). Since this condition is not satisfied by the solution given
in eq.~(\ref{3:PsiConf}), one might consider the possibility that local
scale-invariance only holds for sufficiently large arguments, 
say $\rho\geq \eps$ and to write
\BEQ
\Psi(\rho) \simeq \left\{ \begin{array}{ll} 
\Psi_0 \rho^{x-\wit{x}_2+2\xi-2\wit{\xi}_2} & 
\mbox{\rm ~~;~ if $\rho\leq\eps$}\\
\Psi_{\rm LSI}(\rho) & \mbox{\rm ~~;~ if $\rho\geq\eps$}
\end{array}\right.
\EEQ
where $\Psi_{\rm LSI}(\rho)$ is the expression given by eq.~(\ref{3:PsiConf})
and the constant $\eps$ sets the scale which separates the two regimes.
Finally, the constant $\Psi_0$ is found from the condition that $\Psi(\rho)$
is continuous at $\rho=\eps$. 

In this way, one may obtain an explicit scaling function which at least
satisfies the basic boundary conditions required for a physically sensible
scaling function. For models with an underlying free-field theory, the
approximation eq.~(\ref{8:gl:2}) becomes exact and leads to eq.~(\ref{8:gl:3}),
in agreement with the available exact solutions, 
as we shall review in section~4. 
For the case of phase-ordering, a straightforward, but just a little lengthy 
calculation (see the appendix for details) leads to 
\BEA
\hspace{-2.2truecm}\lefteqn{
\vec{\Psi}(w) = A \Gamma(\lambda_C-2a')\,
{_2F_1}(2\lambda_C-d-2a',\lambda_C-2a';\lambda_C+1-2a'-d/2;-1/w)
w^{2a'-\lambda_C} 
}
\nonumber \\
\hspace{-1.5truecm}& & + B \Gamma(d/2)\,
{_2F_1}(\lambda_C-d/2,d/2;1+2a'+d/2-\lambda_C;-1/w)
w^{-d/2}
\nonumber \\
\hspace{-1.5truecm}& & + A E^{1-2a'}
\frac{2\lambda_C-d-2a'}{\lambda_C+1-2a'-d/2}
\left[ (Ew)^{2a'-1}\gamma(\lambda_C+1-2a',Ew) -\gamma(\lambda_C,Ew)\right]
w^{-\lambda_C}
\nonumber \\
\hspace{-1.5truecm}& & + A E^{-2a'} w^{-\lambda_C}\gamma(\lambda_C,Ew) 
- A\gamma(\lambda_C-2a',Ew)w^{2a'-\lambda_C}
\nonumber \\
\hspace{-1.5truecm}& & + BE^{d/2-\lambda_C} w^{-\lambda_C}\gamma(\lambda_C,Ew) 
- B \gamma(d/2,Ew) w^{-d/2}
\label{gl:PSIFinal} 
\EEA
where ${}_2F_1$ is a hypergeometric function and $\gamma(a,z)$ an incomplete 
gamma-function \cite{Abra65}, 
$E={\cal M}\eps$ and $A,B$ are constants related to $\psi_{0,1}$.
The first two lines would give the scaling function $\vec{\Psi}$ if the 
above-described procedure were unnecessary. For free fields, one has $\lambda_C=d/2$. Setting $A=0$, we then have 
$\vec{\Psi}(w)=B \Gamma(d/2) w^{-d/2}$
and recover the case $T<T_c$ of (\ref{8:gl:2}). 

Together, eqs.~(\ref{3:fCpo},\ref{gl:PSIFinal}) give the final autocorrelation
function $C(t,s)$ of phase-ordering kinetics. 
This is the central result of this section (the special case 
$a=a'$ was announced in \cite{Henk04b}). 

We shall not give here the analogous expression for quenches to $T=T_c$, since
the known exactly solvable cases with $z=2$ (such as the spherical model or
the XY model in spin-wave approximation) are described by a free-field
theory and in more general cases the dynamical 
exponent $z\ne 2$ \cite{Cala05} which in full generality is beyond the 
scope of this article.

\section{Tests}

We now compare the predictions of local scale-invariance -- 
either eq.~(\ref{8:gl:3}) for models with an underlying  free-field
theory or else eqs.~(\ref{3:fCpo},\ref{gl:PSIFinal}) or 
(\ref{3:fCcrit},\ref{gl:PSIFinal}) if that is not the case --
to explicit model results. In table~\ref{tb:results_fun} we collect results
for $f_C(y)$ of some analytically solved models which we now proceed to analyse. 

\begin{table} 
  \[
  \hspace*{-2.5cm}
  \begin{array}{||c|c|c||c|c||}  \hline \hline
    \multicolumn{3}{||c||}{} &  z & f_C(y) \\
    \hline
    \hline
    \multicolumn{3}{||c||}{\mbox{$1D$ Glauber-Ising, $T=0$}} &
    2  & (2/\pi)\arctan( \sqrt{2/(y-1)} ) \\ \hline \hline
\multicolumn{3}{||c||}{\mbox{spherical model, $T<T_c$}} &
    2  & ( (y+1)^2/4y )^{-d/4} \\ \hline \hline
\multicolumn{2}{||c||}{\mbox{spherical, $T=T_c$}} & 2<d<4 & 
    2  & (y-1)^{-\frac{d}{2}+1} y^{1-d/4}(y+1)^{-1} \\ \hline \hline
\multicolumn{2}{||c||}{\mbox{spherical, $T=T_c$}} & d>4 & 
    2  & (y-1)^{-\frac{d}{2}+1}-(y+1)^{-\frac{d}{2}+1} \\ \hline \hline
\multicolumn{3}{||c||}{\mbox{contact process {\sc bcp} = {\sc ew}1}} &
    2  & (y-1)^{-\frac{d}{2}+1} -
    (y+1)^{-\frac{d}{2}+1} \\ \hline \hline
    \mbox{pair} & \alpha < \alpha_C & d > 2 & 2 &
    (y-1)^{-\frac{d}{2}+1} - (y+1)^{-\frac{d}{2}+1}  \\ \cline{2-5}
    \mbox{contact} &  & 2 < d < 4 & 2 & 
    (y+1)^{-\frac{d}{2}} {_2F_1}
    \left(\frac{d}{2},\frac{d}{2};\frac{d}{2}+1;\frac{2}{y+1}\right)\\
    \cline{3-5}
    \mbox{process} & \raisebox{1.6ex}[-1.6ex]{$\alpha =
    \alpha_C$} & d > 4 & 2 &  (y+1)^{-\frac{d}{2}+2}
    - (y-1)^{-\frac{d}{2}+2} + ( d-4 ) (y-1)^{-\frac{d}{2}+1} 
\\ \hline \hline
\multicolumn{3}{||c||}{\mbox{Edwards-Wilkinson {\sc ew}2}} &
    2  & (y-1)^{-\frac{d}{2}+1+\rho} -
    (y+1)^{-\frac{d}{2}+1+\rho} \\ \hline 
\multicolumn{3}{||c||}{\mbox{Mullins-Herring {\sc mh}1}} &
    4  & (y-1)^{-\frac{d}{4}+1} -
    (y+1)^{-\frac{d}{4}+1} \\ \hline
\multicolumn{3}{||c||}{\mbox{Mullins-Herring {\sc mh}2}} &
    4  & (y-1)^{-\frac{d}{4}+1+\rho/2} -
    (y+1)^{-\frac{d}{4}+1+\rho/2} \\ \hline 
\multicolumn{3}{||c||}{\mbox{Mullins-Herring {\sc mh}c}} &
    4  & (y-1)^{-{(d-2)}/{4}} -
    (y+1)^{-{(d-2)}/{4}} \\ \hline \hline
  \end{array}
  \]
\caption[Scaling functions]{Dynamical exponents and scaling functions 
(up to normalization) of the autocorrelator of some analytically solved models, see text for the definitions.}
\label{tb:results_fun}
\end{table}

\subsection{One-dimensional Glauber-Ising model}

Our first example is the $1D$ Ising model with Glauber dynamics \cite{Glau63}, 
at zero temperature and with nearest-neighbour hamiltonian
${\cal H}=-J\sum_i \sigma_i \sigma_{i+1}$, $\sigma_i=\pm 1$. 
For a totally disordered initial state, the exact two-time
autocorrelation and autoresponse functions are \cite{Godr00a,Lipp00}
\BEQ
C(t,s) = \frac{2}{\pi} \arctan\sqrt{\frac{2}{t/s-1}\,} \;\;,\;\;
R(t,s) = \frac{1}{\pi} \sqrt{\frac{1}{2s(t-s)}\,} 
\EEQ
These results remain valid for long-ranged initial conditions
$\langle\sigma_r \sigma_0\rangle\sim r^{-\nu}$ with $\nu>0$ \cite{Henk03d}. 

Comparing $R(t,s)$ with the LSI-prediction eqs.~(\ref{gl:sR},\ref{1:R}) 
one reads off $a=0, a'=-1/2$ and $\lambda_R/z=1/2$ \cite{Pico04}. 
Since the autocorrelation function does not depend on the range of the
initial condition,
it is reasonable to consider $T=0$ as the critical point of the $1D$ 
Ising model. Since $b=a=0$ and $\lambda_C=\lambda_R=1$ and the $1D$ Glauber
model is described in terms of free fermions, we may use (\ref{8:gl:2}) at
$T=T_c=0$ and then have from eq.~(\ref{8:gl:3})
\BEQ
C(t,s) = C_0 \int_0^1 \!\D v\: v^{2\mu} 
\left[\left(\frac{t}{s}-1\right)(1-v)\right]^{-2\mu-1/2} 
\left(\frac{t}{s}+1-2v\right)^{2\mu}
\EEQ
Choosing $\mu=-1/4$ and $C_0=\sqrt{2}/\pi$, the exact result quoted
above is reproduced \cite{Henk06a}. 

\subsection{XY model in spin-wave approximation} 

Consider a system with planar spins $\vec{S}\in\mathbb{R}^2$ and with
$|\vec{S}|=1$. In the spin-wave approximation, the usual nearest-neighbour
hamiltonian becomes
\BEQ
{\cal H} = -\sum_{(\vec{r},\vec{r}')} \vec{S}(\vec{r})\cdot \vec{S}(\vec{r}')
\simeq  \frac{1}{2} \int_{\mathbb{R}^d} \!\D\vec{r}\: (\nabla \phi(\vec{r}))^2
\EEQ
Since the spin-wave approximation considers the small deviations from a
fully ordered state, LSI as formulated here with a vanishing initial
order-parameter is not applicable. Therefore, a comparison of the
existing results \cite{Bert01} 
for the {\em spin} correlators and responses with LSI would
require the inclusion of a non-vanishing initial magnetization into the latter. 
On the other hand, since angular
variable $\phi$ satisfies a simple noisy diffusion equation and has
a vanishing initial value. The exactly known {\em angular}
autoresponse and autocorrelation functions do agree, respectively, with 
eqs.~(\ref{gl:sR},\ref{1:R}) and eq.~(\ref{8:gl:3}) in the critical case $T=T_c$,
where $a=b=d/2-1$ and $\lambda_C=\lambda_R=d$ (for $d>2$)
\cite{Pico04}.\footnote{If $d=2$, $\lambda_R=2$ and $\lambda_C=0$.} 

\subsection{Spherical model}

The spherical model is described by a set of continuous spin variables 
$\phi\in\mathbb{R}$ subject to the spherical constraint $\sum_{\vec{r}}
\langle \phi(t,\vec{r})^2\rangle = {\cal N}$ where $\cal N$ is the number 
of sites of the lattice. In the continuum limit, the equation of motion
is \cite{Godr00b}
\BEQ
\partial_t \phi(t,\vec{r}) = \Delta \phi(t,\vec{r}) -v(t) \phi(t,\vec{r}) 
+\eta(t,\vec{r})
\EEQ
where $\eta$ is the usual gaussian noise with zero mean and variance
$\eta(t,\vec{r})\eta(t,\vec{r}')\rangle=2T\delta(t-t')\delta(\vec{r}-\vec{r'})$.
The time-dependent potential $v(t)$ arises from the spherical constraint. It
can be found from the function $g(t)=\exp(2\int_0^t\!\D\tau\, v(\tau))$ where
$g(t)$ solves the Volterra integral equation, 
for short-ranged initial correlations
\BEQ
g(t) = f(t) + 2T \int_0^t \!\D t'\: f(t-t') g(t') \;\; , \;\;
f(t) = \Theta(t) \left( e^{-4t} I_0(4 t)\right)^d
\EEQ
and $I_0$ is a modified Bessel function \cite{Abra65}. This is exactly
the kind of equations of motion of the type (\ref{2:LangevinE}) 
considered in section~3. Indeed, an explicit
analysis shows \cite{Godr00b,Anni06,Newm90} that $g(t)\sim t^{\digamma}$, 
with $\digamma=d/4$ if $T<T_c$. If $T=T_c$, then $\digamma=1-d/4$ for
$2<d<4$ and $\digamma=0$ if $d>4$. It is therefore clear that the predictions
(\ref{1:R}) and (\ref{8:gl:3}) of local scale-invariance will be satisfied
{\em both} below and at criticality (since $z=2$ throughout)
as indeed has been checked explicitly \cite{Pico04}, see also 
table~\ref{tb:results_fun}.

\subsection{Two-dimensional Ising and Potts models} 

All the exactly solved models reviewed so far are special in that the
response and correlation  functions satisfy linear equations of motion and
therefore only test the rather weak prediction (\ref{8:gl:3}). It
is therefore significant that there is evidence in favour of LSI (including
the non-conventional conformal invariance described in section~3.2) from models
whose equations of motion are  non-linear. We refer to 
\cite{Henk07a,Henk07b} for the response function and concentrate on the
correlation function for $T<T_c$. 

\begin{table}
\begin{center}
\begin{tabular}{|c|l|llll|cl|} \hline 
$q$ & $T/T_c$ & $\lambda_C$ & $A$ & $B$ & $E$ & update & Ref. \\ \hline
2   & 0       & 1.25(1) & -0.601 & 3.94 & 0.517 & {\sc r} & \cite{Henk04b} \\
    & 0.6610  & 1.25(1) & -5.41  & 18.4 & 1.24  & {\sc r} & \cite{Henk04b} \\
    & 0.6610  & 1.24(2) & -5.41  & 18.4 & 1.24  & {\sc c} & \cite{Lore07}\\\hline
3   & 0.5000  & 1.19(3) & -0.05  & 2.15 & 0.6   & {\sc c} & \cite{Lore07}\\\hline
8   & 0.5001  & 1.25(1) & -0.07  & 1.98 & 0.4   & {\sc c} & \cite{Lore07}\\\hline
\end{tabular} \end{center}
\caption{Parameters describing the autocorrelation function of $2D$ Ising and
$q$-states Potts models, according to LSI, eq.~(\ref{gl:PSIFinal}). The letters
{\sc r} and {\sc c} refer to random and checkerboard update, respectively.  
\label{tab:ABE}}
\end{table}

To this end, the two-time autocorrelation function was calculated in the
$2D$ kinetic Ising model with heat-bath dynamics \cite{Henk04b} as well as
for the $q$-states Potts model, with $q=2,3,8$ \cite{Lore07,Jank06,Lore06}. 
The two-time autocorrelation was measured as
\BEQ
\hspace{-2.0truecm}C(t,s)_{\rm Ising} = \frac{1}{\cal N} 
\sum_i\langle\sigma_i(t)\sigma_i(s)\rangle \;\; , \;\;
C(t,s)_{\rm Potts} =\frac{1}{q-1}\left(\frac{q}{\cal N}\sum_i \langle
\delta_{\sigma_i(t),\sigma_i(s)} \rangle -1 \right)
\EEQ
where the sums run over the entire lattice with $\cal N$ sites. In 
the first case, the $\sigma_i=\pm 1$ are usual Ising spins, whereas
in the second case $\sigma_i=1,2,\ldots, q$ are Potts spins. Another difference
between the two simulations is that either random or checkerboard update
was used. The results for the parameters describing $C(t,s)$ from the
two simulations are shown in table~\ref{tab:ABE}
and allow some explicit tests of universality. Throughout, the relation
$a=a'$ was assumed. First, the estimates for the 
exponent $\lambda_C$ in the $q=2$ universality class, in agreement with earlier results \cite{Fish88,Brow97}, appear to be independent of the
lattice realization and of the temperature, as long $T<T_c$, which is
of course expected from renormalization arguments \cite{Bray94}. 
It is also satisfying to see that the same values of the parameters $A,B,E$ of
the LSI formula (\ref{3:fCpo},\ref{gl:PSIFinal}) can be used for two
different update schemes. One observes that the term parametrized by $B$ is 
apparently much more important than the other one, but the ratio $A/B$ 
apparently depends on temperature. 

\begin{figure}
\begin{center}
\includegraphics[angle=0,width=0.51\textwidth]{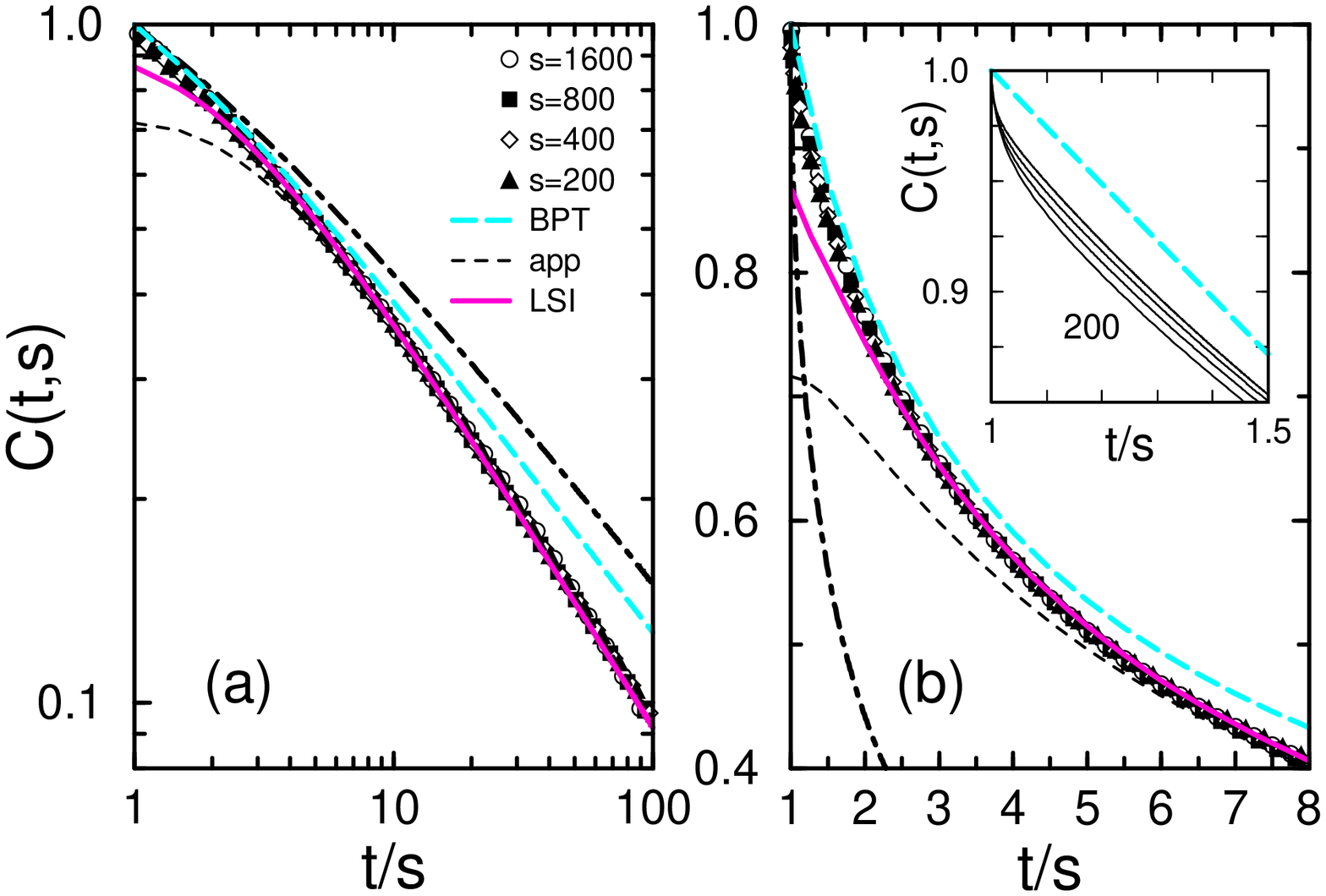}\hspace*{2mm}%
\includegraphics[angle=0,width=0.51\textwidth]{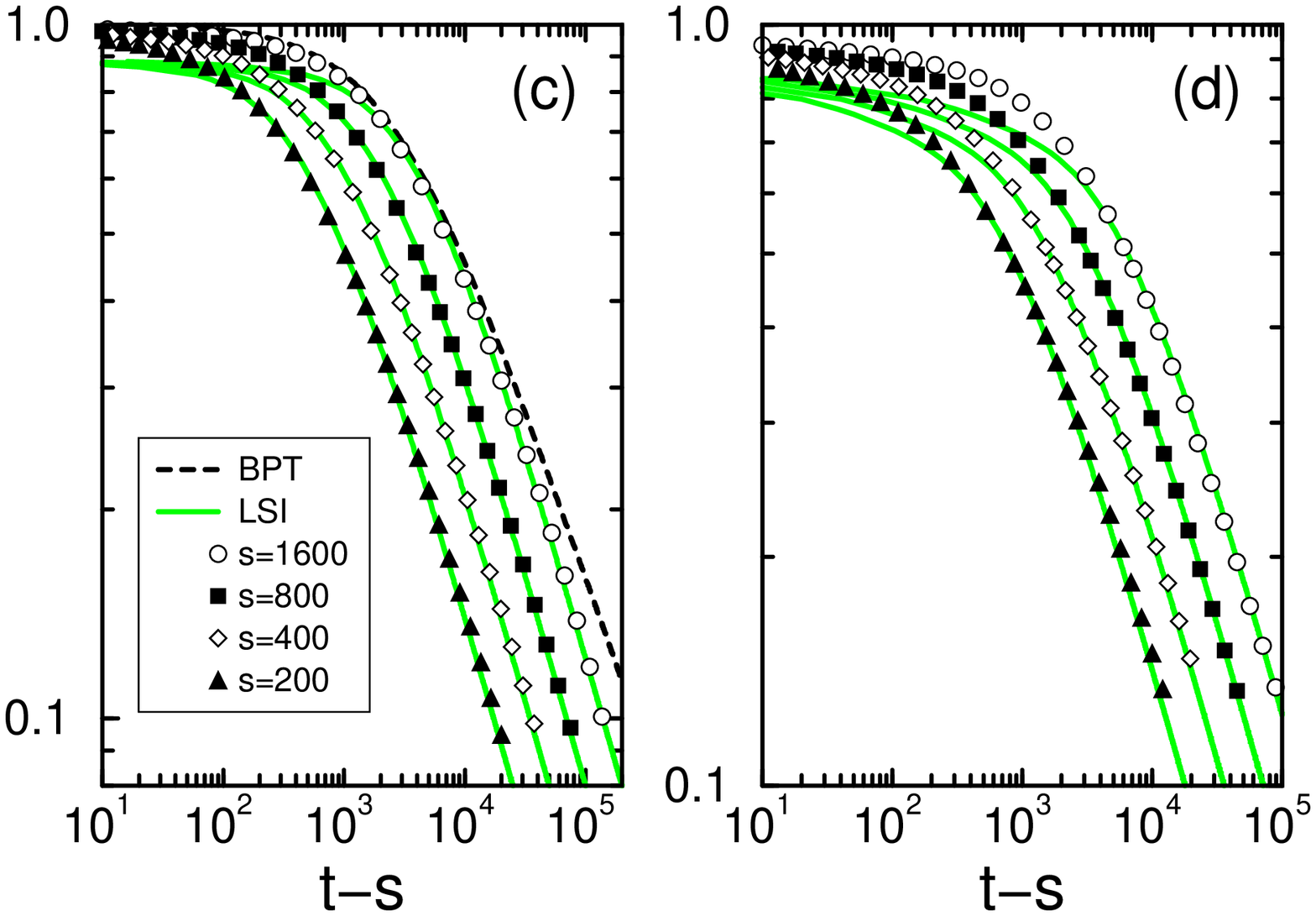}
\caption[a]{\label{fig:C(t,s)Ising} Two-time autocorrelation function $C(t,s)$ 
for the 2D Ising model and comparison with the LSI prediction
(\ref{3:fCpo},\ref{gl:PSIFinal}). In (a) the scaling behaviour for $T=0$ 
is shown for a large range of values of $y=t/s$  and (b) redisplays the same data for smaller values of $y$. The inset compares the data for $y\approx 1$ and
$s=[200,400,800,1600]$ with the closed approximation of BPT \cite{Bray91,Toyo92}. 
The dash-dotted line in (a) is the second-order calculation 
from \cite{Maze98} and the dash-dotted line in (b) gives the result from \cite{Liu91}, see text. The autocorrelation is re-plotted 
as a function of $t-s$ and compared to LSI for $T=0$ in (c) and for $T=1.5$ 
in (d). \cite{Henk04b}.}
\end{center}
\end{figure}

In figures \ref{fig:C(t,s)Ising} and \ref{fig:C(t,s)} the degree of agreement
between the numerical data and the prediction (\ref{3:fCpo},\ref{gl:PSIFinal}) 
is shown. Clearly, the system ages, as can be seen in 
figure~\ref{fig:C(t,s)Ising}cd, and there is excellent dynamical scaling for all
models considered. For the Ising model, we show in figure~\ref{fig:C(t,s)Ising}ab
the free-field approximation (\ref{8:gl:3}), labelled `app', 
which only describes the data for rather large values of $y=t/s$. This is
hardly surprising, since the model is known not to be described by a free field.
On the other hand, we see that in all
models the LSI predictions eqs.~(\ref{3:fCpo},\ref{gl:PSIFinal}) give a good
overall description over the whole range of the scaling variable, with the
only exception of the relatively small region $t/s\lesssim 2-3$. It is
conceivable that the patching procedure outlined in section~3 is not precise 
enough. It remains an open problem to derive a quantitatively precise formula
which reproduces the numerical data over the entire range $1<y<\infty$. 

\begin{figure}
\begin{center}
\includegraphics[angle=-90,width=0.32\textwidth]{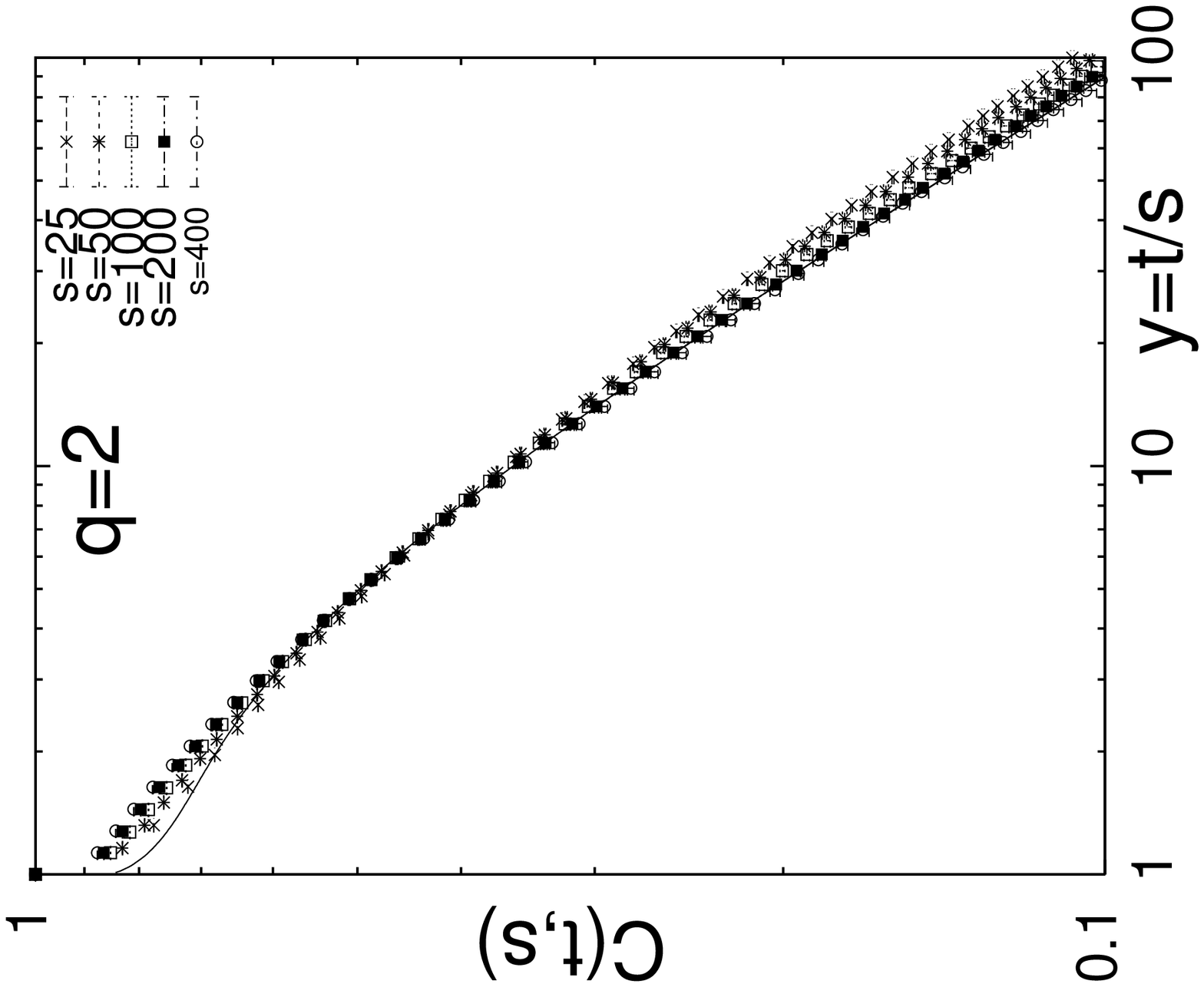}\hspace*{3mm}%
\includegraphics[angle=-90,width=0.32\textwidth]{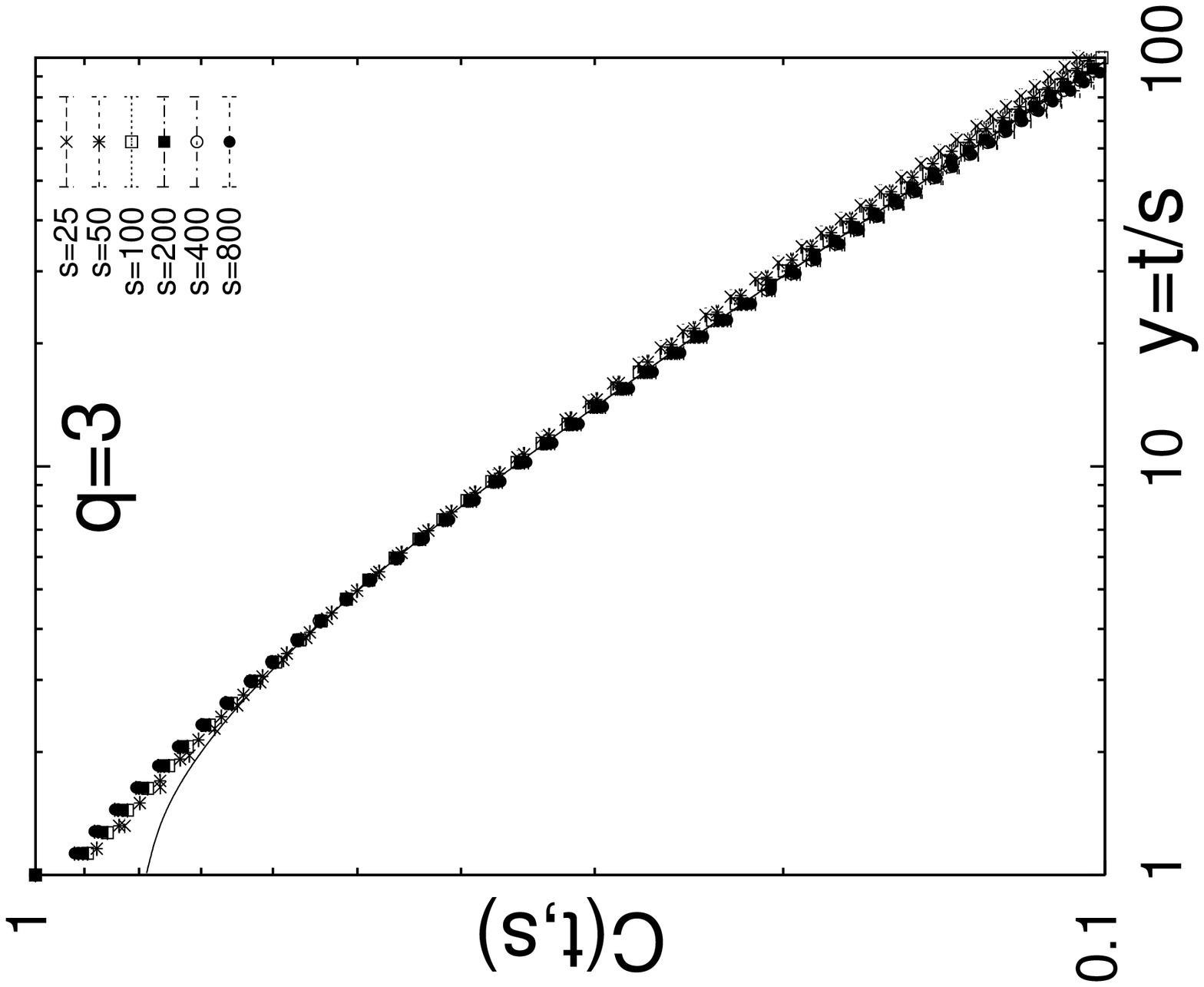}\hspace*{3mm}%
\includegraphics[angle=-90,width=0.32\textwidth]{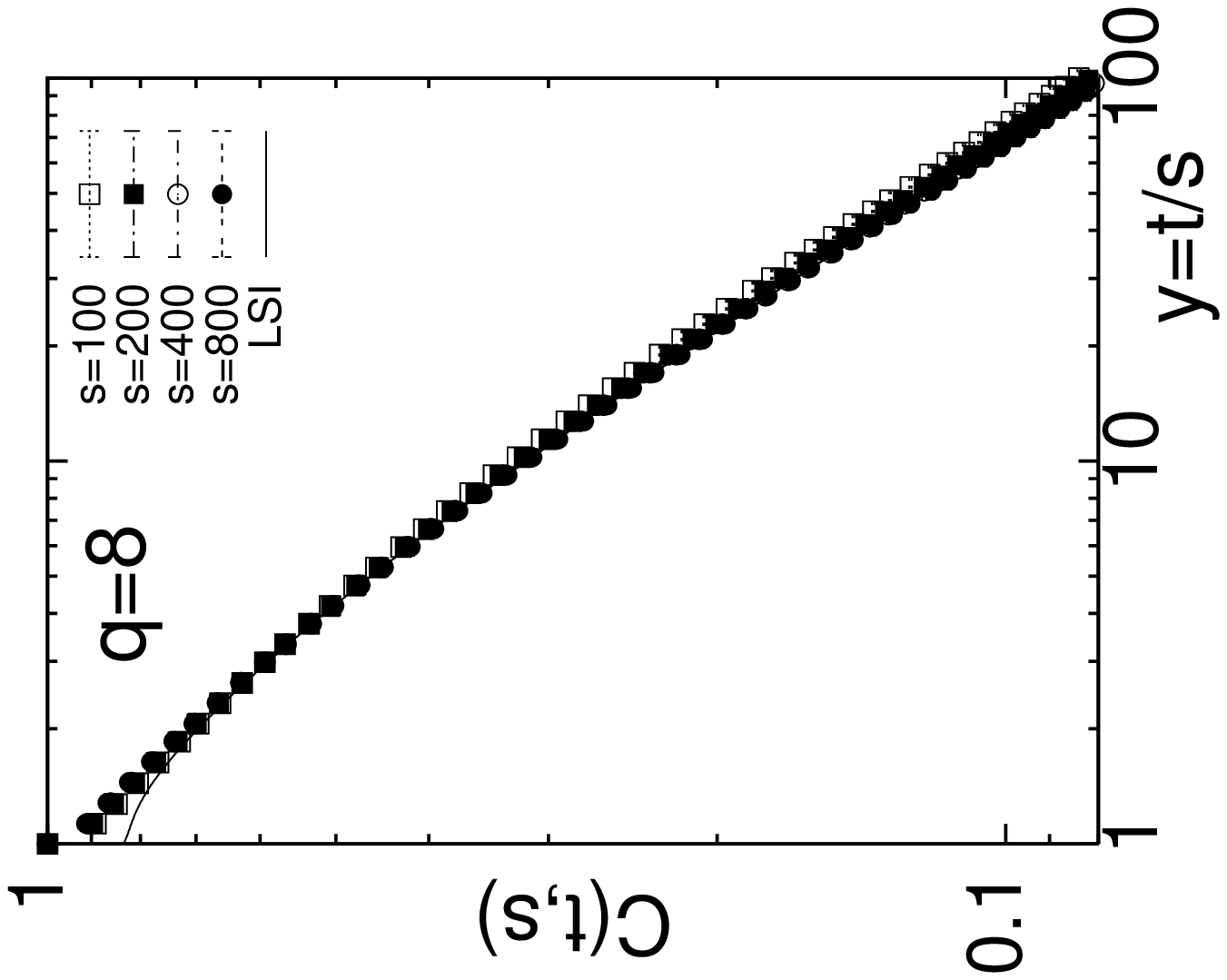}
\caption[a]{\label{fig:C(t,s)} Two-time autocorrelation function for the $2D$
$q$-states Potts model with $q=2$ (left), $q=3$ (middle) and $q=8$ (right) 
\cite{Lore07,Jank06}. 
The different data symbols correspond to different waiting times. 
The solid lines show the fits to the scaling prediction (\ref{gl:PSIFinal})
based on local scale invariance.}
\end{center}
\end{figure}

On the other hand, the difficulty of achieving any quantitative agreement 
of an analytical approach and the numerical data is illustrated by comparing
with the results of earlier attempts to find $C(t,s)$ in phase-ordering 
kinetics. Using Gaussian closure procedures, Bray, Puri and Toyoki (BPT) 
find for the O($n$)-model \cite{Bray91,Toyo92,Roja99}
\BD
\hspace{-2.2truecm}f_{C,{\rm BPT}}(y) = \frac{n}{2\pi} \left[ B\left(\frac{1}{2},\frac{n+1}{2}\right)\right]^2
\left(\frac{4y}{(y+1)^2}\right)^{d/4}\:
{_2F_1}\left(\frac{1}{2},\frac{1}{2};\frac{n+2}{2};
\left(\frac{4y}{(y+1)^2}\right)^{d/2}\right)~
\ED
where $B$ is Euler's beta function \cite{Abra65}. While this form does follow
the general trend of the data, especially if $y$ is not too large, but without
reproducing them entirely, it mainly suffers
from the disadvantage that it predicts the autocorrelation exponent $\lambda_C=d/2$ in disagreement with the data, see also 
figure~\ref{fig:C(t,s)Ising}c. Liu and Mazenko \cite{Liu91}
and Mazenko \cite{Maze98} tried to remedy this by constructing more elaborate
analytical schemes of which the BPT formula represents the lowest order. 
In particular, this leads for the $2D$ Ising model to fairly accurate values 
$\lambda_C=1.246(20)$ \cite{Liu91} and $\lambda_C=1.105\ldots$ \cite{Maze98}
(Mazenko also showed \cite{Maze04} that $\lambda_R=\lambda_C$ in his scheme). 
However, the detailed behaviour for $f_C(y)$ deviates largely from the
data, as can be seen from the dash-dotted lines in figure~\ref{fig:C(t,s)Ising}b
for the prediction of \cite{Liu91} -- in agreement with the earlier tests performed by Brown {\it et al.} \cite{Brow97} -- and in figure~\ref{fig:C(t,s)Ising}a for the prediction of \cite{Maze98}. These
examples illustrate the extreme difficulty of describing $C(t,s)$ theoretically. 
We consider it remarkable that local scale-invariance achieves, for the first 
time, a precise representation of the numerical data over almost the entire range
of the scaling variable $y$. 

A similar behaviour as for the Ising model is found for the Potts models 
with $q=3$ as well as $q=8$ \cite{Lore07}. 
This is the first time that the ageing behaviour and LSI of the autocorrelator
was tested for a system which undergoes a first-order transition. 

The examples presented here are the only ones which presently allow to test
the extension $\mathfrak{age}_d \to \mathfrak{sch}_d \to \mathfrak{conf}_{d+2}$ 
of the local dynamical symmetry.

\subsection{Bosonic contact and pair-contact processes}

\begin{figure}[htb] 
  \vspace{0.5cm}
  \centerline{\epsfxsize=5.0in\epsfclipon\epsfbox
  {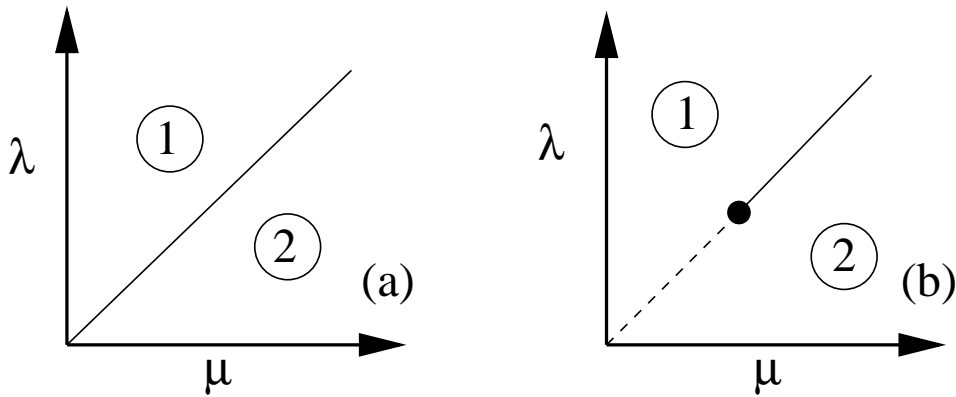}
  }
  \caption[Phase diagramme of the bosonic contact and pair-contact 
  processes]{Schematic  
  phase-diagrammes for $D\ne 0$ of (a) 
  the bosonic contact process and the bosonic pair-contact process in $d\leq 2$ 
  dimensions and (b)
  the bosonic pair-contact process in $d>2$ dimensions. The absorbing region 1, 
  where $\lim_{t \rightarrow \infty} \rho(\vec{x},t)= 0$, is separated
  by the critical line eq.~(\ref{gl:4:bcp:Krit}) from the active region 2, 
  where $\rho(\vec{x},t)\to\infty$ as $t\to\infty$. Clustering along
  the critical line is indicated in (a) and (b) by full lines, 
  but in the bosonic pair-contact process with $d>2$ the steady-state may also 
  be homogeneous (broken line in (b)). These two regimes are separated by a
  multicritical point. 
  \label{abb:bcp:Abb0}}
\end{figure}

Two exactly solvable models allow to test the of LSI in 
situations when the stationary states are no longer equilibrium states,
which requires a modification of the reduction formul{\ae} to the
deterministic part. See \cite{Henk07a} and references therein for the
motivation of these models, which are defined 
as follows \cite{Houc02,Paes04,Baum05a}. 
Consider a set of particles of a single species 
$A$ which move on the sites of a $d$-dimensional hyper-cubic lattice. 
On any site one may have an arbitrary (non-negative) number of particles.
Single particles may hop to a nearest-neighbour site with unit rate and in 
addition, the following single-site creation and annihilation processes 
are admitted 
\BEQ \label{gl:4:bcp:rates}
m A \stackrel{\mu}{\longrightarrow} (m+1) A \;\; , \;\; 
p A \stackrel{\lambda}{\longrightarrow} (p-\ell) A \;\; ; \;\;
\mbox{\rm with rates $\mu$ and $\lambda$} 
\EEQ
where $\ell$ is a positive integer such that $|\ell| \leq p$. 
Consider the following special cases:
\begin{enumerate}
\item {\em critical bosonic contact process:} ({\sc bcp}) 
$p=m=1$. Here only $\ell = 1$ is possible. Furthermore
the creation and annihilation rates are set equal $\mu =\lambda$.
\item {\em critical bosonic pair-contact 
process:} ({\sc bpcp}) $p=m=2$. We fix
$\ell=2$, set $2 \lambda = \mu$ and define the control parameter
\footnote{For a coagulation process $2A\to A$, 
where $\ell=1$, analogous results hold true if
one sets $\lambda = \mu$ and $\alpha = {\mu}/{D}$.}
\BEQ
\alpha := \frac{3 \mu}{2 D}
\EEQ
\end{enumerate}
Linear equations of motion for one- and two-point functions may be derived
straightforwardly \cite{Houc02,Paes04,Baum05a} and it follows that in both
models the spatial average of the local particle-density 
$\int\!\D\vec{r}\, \rho(t,\vec{r})=\rho_0$ remains constant in time. Scaling
occurs along the critical line \cite{Paes04} 
\BEQ \label{gl:4:bcp:Krit}
\ell \lambda = \mu.
\EEQ
The physical behaviour of these models is indicated in the phase-diagramme
figure~\ref{abb:bcp:Abb0}. Whereas in the {\sc bcp} the variance 
$\langle\rho^2(t,\vec{r})\rangle$ diverges
as $t\to\infty$ for dimensions $d\leq 2$ but stays finite for $d>2$ such that
one has the same behaviour along the critical line, 
in the {\sc bpcp} there is a critical value $\alpha_C$ of the control 
parameter, given by
\BEQ
\frac{1}{\alpha_C} = 2 \int_0^{\infty} \!\D u\, \left( e^{-4u} I_0(4u)\right)^d
\label{gl:def_alphaC}
\EEQ
Specific values are $\alpha_C(3) \approx 3.99$, $\alpha_C(4) \approx 6.45$ 
and $\lim_{d\,\searrow\, 2}\alpha_C(d)=0$. Now the variance 
$\langle\rho^2(t,\vec{r})\rangle$ stays finite on the critical line for
$\alpha<\alpha_C$ but diverges otherwise. 

This means that at the multicritical point at $\alpha=\alpha_C$ there occurs a 
{\em clustering transition} such that for $\alpha<\alpha_C$ the systems evolves
towards a more or less homogeneous state while for 
$\alpha\geq \alpha_C$ particles
accumulate on very few lattice sites while the other ones remain empty. 
In contrast with the bosonic contact process, clustering occurs in some region
of the parameter space for all values of $d$. 

Solving the equations of motion with uncorrelated initial conditions 
yields the exact two-time autocorrelator\footnote{In the critical linear 
voter model \cite{deOl93,Sast03}, the exact two-time autocorrelator 
\cite{Dorn98} agrees in $1D$ with the one of the $1D$ Glauber-Ising model 
and for $d>2$ with the one of the {\sc bcp}.}
(see table~\ref{tb:results_fun} and note that there is no scaling for
$\alpha>\alpha_C$) which may also be written in an
integral form 
\BEQ
\label{gl:4:bcp:formaintegralC}
f_C(y) = \mathcal{C}_0 \int_0^1 \!\D\theta\, 
\theta^{a-b} (y+1-2 \theta)^{-{d}/{2}}
\EEQ
quite similar to (\ref{8:gl:3}). 
While the relation $a=b$ still holds true for the {\sc bcp} and the 
{\sc bpcp} with $\alpha<\alpha_C$, that is no longer true for the
{\sc bpcp} at its multicritical point $\alpha=\alpha_C$ \cite{Baum05a}.  

We now review how the formalism of LSI can be adapted to these models, 
following \cite{Baum05b}.

\subsubsection{Bosonic contact process}

{}From a field-theoretic point of view, the {\sc bcp} can be 
described in terms of an order-parameter field $\phi$ and a response
field $\wit{\phi}$, defined  such that 
$\langle \phi(t,\vec{r})\rangle =\langle \wit{\phi}(t,\vec{r})\rangle = 0$. 
The action \cite{Howa97,Taeu05,Taeu06} is again decomposed
${\cal J}[\phi,\wit{\phi}]={\cal J}_0[\phi,\wit{\phi}]+
{\cal J}_b[\phi,\wit{\phi}]$ into
a `deterministic' part 
\BEQ
{\cal J}_0[\phi,\wit{\phi}] := \int_{\mathbb{R}^d} \!\D \vec{R} \int_0^{\infty} 
\!\D u\: \left[ \wit{\phi} (2\mathcal{M}\partial_u - \nabla^2)\phi \right]
\EEQ
and which is manifestly Galilei-invariant, 
whereas the `noise' now contains two contributions and is described by 
\BEQ
{\cal J}_b[\phi,\wit{\phi}] := - \mu \int_{\mathbb{R}^d} \!\D \vec{R} 
\int_0^{\infty} \!\D u\: 
\left[{\wit{\phi}}^2 (\phi+\rho_0) \right].
\EEQ
where uncorrelated initial conditions are implied.

In what follows, some composite fields will be needed. 
Together with their scaling dimensions and their masses, they are listed in 
table~\ref{tab:champs}. 
\begin{table}
\begin{center}
\begin{tabular}{|c|cr|} \hline 
field & scaling dimension & mass \\ \hline 
$\phi$       & $x$ & $\cal M$ \\
$\wit{\phi}$ & $\wit{x}$ & $-{\cal M}$ \\
${\wit{\phi}}^2$ & $\wit{x}_2$ & $-2{\cal M}$ \\
$\Upsilon := {\wit{\phi}}^2 \phi$ & $x_{\Upsilon}$ & $-{\cal M}$ \\ 
$\Sigma :=  {\wit{\phi}}^3 \phi$ & $x_{\Sigma}$ & $-2 {\cal M}$  \\ 
$\Gamma:=  {\wit{\phi}}^3 \phi^2$ & $x_{\Gamma}$ & $-{\cal M}$  \\ \hline
\end{tabular} \end{center}
\caption{Scaling dimensions and masses of some composite fields.
\label{tab:champs}}
\end{table}
For free fields one has
$\wit{x}_2 = 2 \wit{x}$, $x_{\Upsilon} = 2 \wit{x}+x$, 
$x_{\Sigma} = 3\wit{x} + x$ and $x_{\Gamma} = 3\wit{x} + 2 x$, 
but not necessarily so for interacting fields. On the other
hand, from the Bargman superselection rules we
expect that the masses of the composite fields as given in
table~\ref{tab:champs} should remain valid for interacting fields as well.

As for magnets, the response function can reduced to a form which does not
contain the noise explicitly \cite{Baum05b}. The correlator becomes 
\BEA
\hspace{-2.0truecm}C(t,s;\vec{r}-\vec{r}') &= &\left\langle \phi(t,\vec{r})
\phi(s,\vec{r}') \exp \left( -\mu \int \!\D \vec{R} \int \!\D u\:
\wit{\phi}^2(u,\vec{R}) \phi(u,\vec{R}) \right) \right.
\nonumber \\ 
\hspace{-2.0truecm}& & \left. ~~ \times 
\exp \left( -\mu \rho_0 \int \!\D \vec{R} \int \!\D u\:
\wit{\phi}^2(\vec{R},u) \right) \right\rangle_0
\nonumber \\
\hspace{-2.0truecm}&=& - \mu \rho_0 \int \!\D \vec{R}
\int \!\D u\: \left\langle \phi(t,\vec{r}) \phi(s,\vec{r}')
\wit{\phi}^2(u,\vec{R}) \right\rangle_0
\\
\hspace{-2.0truecm}& & + \frac{\mu^2}{2} \int \!\D
\vec{R} \D \vec{R}' \int \!\D u \D u'\: \left\langle \phi(t,\vec{r})
\phi(s,\vec{r}') \Upsilon(u,\vec{R}) \Upsilon(u',\vec{R}') \right\rangle_0
\nonumber
\EEA
using the field $\Upsilon$, see table~\ref{tab:champs}. 
Hence the connected correlator is determined by three- 
and four-point functions of the noiseless theory. 

The noiseless three-point response can be found
from its covariance under the ageing algebra $\mathfrak{age}_d$ \cite{Baum05b}
\BEA
\langle \phi(t,\vec{r}) \phi(s,\vec{r}') \wit{\phi}^2
(u,\vec{R}) \rangle_0 &=&
(t-s)^{x-\frac{1}{2}\tilde{x}_2} (t-u)^{-\frac{1}{2} \tilde{x}_2}
(s-u)^{-\frac{1}{2} \tilde{x}_2} \nonumber \\ & &
\hspace{-5.5truecm} \times \exp \left( -\frac{\mathcal{M}}{2}
\frac{(\vec{r}-\vec{R})^2}{t-u} - \frac{\mathcal{M}}{2}
\frac{(\vec{r}'-\vec{R})^2}{s - u} \right) \Psi_3 ( u_1,v_1)
\Theta(t-u) \Theta(s-u)
\label{gl:55}
\EEA
with
\BEA
u_1 & = & \frac{u}{t}\cdot \frac{ [(s-u)(\vec{r}-\vec{R})
- (t-u)(\vec{r}' - \vec{R})]^2 }{(t-u) (s-u)^2} \nonumber \\
v_1 & = & \frac{u}{s}\cdot \frac{ [(s-u)(\vec{r}-\vec{R})
- (t-u)(\vec{r}' - \vec{R})]^2 }{(t-u)^2 (s-u)} 
\EEA
and an undetermined scaling function $\Psi_3$. Hence the critical {\sc bcp} 
is described by a free field-theory, one can expect that 
$x=\wit{x}=d/2$, hence $\tilde{x}_2 = d$, $x_{\Upsilon} = \frac{3}{2} d$
for the composite fields. Therefore the autocorrelator takes the general form
$C(t,s) = s^{1-d/2} g_1(t/s) + s^{2-d} g_2(t/s)$ and for $d>2$ the second
term, which comes from the four-point function, merely furnishes a finite-time
correction to scaling. Now, if one chooses in eq.~(\ref{gl:55}) \cite{Pico04}
\BEQ \label{gl:3:Xi}
\Psi_3(u_1,v_1) = \Xi \left(\frac{1}{u_1}-\frac{1}{v_1}\right)
\EEQ
where $\Xi$ remains an arbitrary function, then 
\BEA
C(t,s) &=& -\mu \rho_0 s^{\frac{d}{2}+1-x-\frac{1}{2}\tilde{x}_2}
(t/s-1)^{\frac{1}{2} \tilde{x}_2 - x - \frac{d}{2}} 
\nonumber \\
& & \times \int_0^1 \!\D \theta\,
[(t/s -\theta)(1-\theta)]^{\frac{d}{2}-\frac{1}{2} \tilde{x}_2}
\vec{\Psi}\left(\frac{t/s+1-2 \theta}{t/s-1} \right)
\EEA
where the function $\vec{\Psi}$ is defined by 
\BEQ
\vec{\Psi}(w) = \int_{\mathbb{R}^d} \!\D \vec{R}\, \exp\left(
-\frac{\mathcal{M}w}{2} \vec{R}^2 \right) \Xi(\vec{R}^2)
\EEQ
and the LSI-prediction for the critical point (\ref{8:gl:3}) is indeed
recovered. The specific result for the {\sc bcp} as listed 
in table~\ref{tb:results_fun} follows for 
$\vec{\Psi}(w)=\Psi_0 w^{-1-a}$.

\subsubsection{Bosonic pair-contact process}

In this case, a new ingredient will be needed. 
The action now reads \cite{Howa97}
${\cal J}[\phi,\wit{\phi}] ={\cal J}_0[\phi,\wit{\phi}] + 
{\cal J}_b[\phi,\wit{\phi}]$ with the `deterministic' part  
\BEQ
\label{gl:sigma02}
{\cal J}_0[\phi,\wit{\phi}] := \int_{\mathbb{R}^d} \!\D \vec{R} 
\int_0^{\infty} \!\D t\:
\left[ \wit{\phi} (2\mathcal{M}\partial_t - \nabla^2)\phi - \alpha\wit{\phi}^2
\phi^2\right].
\EEQ
and the noise part
\BEA
\label{gl:kappa2}
\hspace{-1.8cm}
{\cal J}_b[\phi,\wit{\phi}] & = & \int_{\mathbb{R}^d} \!\D \vec{R} 
\int_0^{\infty} \!\D u\:
\left[- \alpha \rho_0^2 \wit{\phi}^2 - 2 \alpha \rho_0\wit{\phi}^2  \phi
-\mu \wit{\phi}^3 \phi^2- 2 \mu \rho_0
\wit{\phi}^3 \phi - \rho_0^2 \wit{\phi}^3 \right]
\EEA
The representations of $\mathfrak{age}_d$ or $\mathfrak{sch}_d$ considered in 
sections~2 and~3 can no longer be used, since the equation of motion
associated to $J_0$ is non-linear, viz. 
\BEQ \label{gl:NLS}
2\mathcal{M}\partial_t \phi(t,\vec{r}) = \nabla^2 \phi(t,\vec{r}) 
-g \phi^2(t,\vec{r}) \wit{\phi}(t,\vec{r})
\EEQ
Rather, new representations of $\mathfrak{age}_d$ and $\mathfrak{sch}_d$
must be constructed which take into account that
$g$ is a {\em dimensionful} quantity which transforms under
local scale-transformations \cite{Stoi05,Baum05b} but the 
generators will not be written down explicitly here for the sake of brevity. 
It can then be shown that 
(i) {\em eq.~(\ref{gl:NLS}) is Schr\"odinger-invariant for any 
value\footnote{If $g$ is taken to be a dimensionless constant, 
it is a well-known mathematical fact that Schr\"odinger-invariance of 
(\ref{gl:NLS}) only holds true in $2D$, see e.g. \cite{Fush93}.} of $d$} 
and (ii) {\em the Bargman superselection rules (\ref{gl:5:Bargman}) 
still apply}. 

Turning to the calculation of the autocorrelation function, one must consider
five possible contributions. It turns out \cite{Baum05b} that the results of 
table~\ref{tb:results_fun} for $f_C(y)$ in the {\sc bpcp}
can be reproduced from the single term 
\BEQ \label{gl:4:G1}
C(t,s) = \alpha \rho_0^2 \int_{\mathbb{R}^d} \!\D \vec{R} 
\int_0^{\infty} \!\D u\: \left\langle
\phi(t,\vec{r}) \phi(s,\vec{r}) \wit{\phi}^2(u,\vec{R})
\right\rangle_0
\EEQ
The required $\mathfrak{age}_d$-invariant three-point function now reads,
where $x=a+1=d/2$ is read off from the response function \cite{Baum05b} 
\BEA
\left\langle \phi(t,\vec{r}) \phi(s,\vec{r}') \tilde{\phi}^2(u,\vec{R})
\right\rangle_0 &=&
(t-s)^{x-\frac{1}{2} \tilde{x}_2} (t-u)^{-\frac{1}{2} \tilde{x}_2}
(s-u)^{-\frac{1}{2} \tilde{x}_2} \nonumber \\ & &
\hspace{-4.0truecm} \times \exp \left( -\frac{\mathcal{M}}{2}
\frac{(\vec{r}-\vec{R})^2}{t-u} - \frac{\mathcal{M}}{2}
\frac{(\vec{r}'-\vec{R})^2}{s - u} \right) \tilde{\Psi}_3 (
u_1,v_1,\beta_1,\beta_2,\beta_3)
\EEA
with
\BEA
u_1 & = & \frac{u}{t}\cdot \frac{ [(s-u)(\vec{r}-\vec{R})
- (t-u)(\vec{r}' - \vec{R})]^2 }{(t-u) (s-u)^2} \\
v_1 & = & \frac{u}{s}\cdot \frac{ [(s-u)(\vec{r}-\vec{R})
- (t-u)(\vec{r}' - \vec{R})]^2 }{(t-u)^2 (s-u)}  \\
\beta_1 & = & \frac{1}{s_2} \cdot \frac{\alpha^{1/y}}{(t-u)^2},
\; \; \beta_2  =  \frac{1}{s_2} \cdot \frac{\alpha^{1/y}}{(s-u)^2},
\;\; \beta_3  =  \alpha^{1/y} s_2 \\
s_2 & = & \frac{1}{t-u} + \frac{1}{u}
\EEA
where $y$ is the scaling dimension of $g$. 
Consider the  following choice for $\tilde{\Psi}_3$ \cite{Baum05b}
\BEQ
\label{gl:some_equation}
\tilde{\Psi}_3 (u_1,v_1,\beta_1,\beta_2,\beta_3) = \Xi \left(
\frac{1}{u_1} - \frac{1}{v_1} \right) \left[ -
\frac{(\sqrt{\beta_1}-\sqrt{\beta_2}) \sqrt{\beta_3}
}{\beta_3 - \sqrt{\beta_2 \beta_3} } \right]^{(a-b)}
\EEQ
where the scaling function $\Xi$ was already given in 
eq.~(\ref{gl:3:Xi}), as for the bosonic contact process. 
We now have to distinguish
the two different cases: first, if $\alpha < \alpha_C$ then $a=b$ \cite{Baum05a}
and we are back to the expressions found for the {\sc bcp}. Second, if 
$\alpha = \alpha_C$ then $a-b=d/2-1$ for $2<d<4$ and $a-b=1$ for $d>4$, 
respectively \cite{Baum05a}. Identifying $\wit{x}_2=2(b-a)+d$, one
finally obtains 
\BEA
C(t,s) &=& s^{-b} (y-1)^{(b-a)-a-1} \int_0^1 \!\D \theta\,
[(t/s-\theta)(1-\theta)]^{a-b}\nonumber \\ 
& & \times \vec{\Psi}\left( \frac{t/s+1-2 \theta}{t/s-1}
\right) \left[ \frac{\theta (t/s-1)}{(t/s-\theta) (1-\theta)}
\right]^{a-b} 
\EEA
The specific results of table~\ref{tb:results_fun} for the {\sc bpcp} 
are recovered if one uses the free-field form 
$\vec{\Psi}(w)=\Psi_{0} w^{-1-a}$.

\section{Growth models and an outlook towards $z\ne 2$}

A different class of non-equilibrium models considers the ballistic 
deposition of particles on a surface. The state of this surface may be 
described in terms of a height variable $h(t,\vec{r})$. For 
irreversible deposition, the system clearly never arrives at an 
equilibrium state. Working in the frame co-moving with the mean 
surface height, the simplest kinetic equation one may write in the case
without mass conservation 
is the well-known Edwards-Wilkinson ({\sc ew}) model \cite{Edwa82}
\BEQ \label{gl:EW}
\partial_t h(t,\vec{r}) = D \nabla^2 h(t,\vec{r}) +\eta(t,\vec{r})
\EEQ
However, if mass conservation must be taken into account, one 
might rather consider the Mullins-Herring ({\sc mh}) model, 
see \cite{Mull63,Wolf90}
\BEQ \label{gl:MH}
\partial_t h(t,\vec{r}) = -D (\nabla^2)^2 h(t,\vec{r}) + \eta(t,\vec{r})
\EEQ
Following \cite{Roet06}, the following types of gaussian noise with
vanishing first moment $\langle \eta(t,\vec{r})\rangle=0$ will be
considered:
\begin{enumerate}
\item[(a)] non-conserved, short-ranged
$\langle \eta(t,\vec{r})\eta(s,\vec{r}')\rangle = 2D
\delta(\vec{r}-\vec{r}')\delta(t-s)$.
\item[(b)] non-conserved, long-ranged
$\langle \eta(t,\vec{r})\eta(s,\vec{r}')\rangle = 2D
|\vec{r}-\vec{r}'|^{2\rho-d}\delta(t-s)$ and $0<\rho<d/2$.
\item[(c)] conserved, short-ranged
$\langle \eta(t,\vec{r})\eta(s,\vec{r}')\rangle = -2D
\nabla_{\vec{r}}^2 \delta(\vec{r}-\vec{r}')\delta(t-s)$.
\end{enumerate}
Then the following models were studied in \cite{Roet06}:
\begin{enumerate}
\item {\sc ew}1: eq.~(\ref{gl:EW}) with the non-conserved noise (a). 
\item {\sc ew}2: eq.~(\ref{gl:EW}) with the non-conserved, long-ranged noise (b).
\item {\sc mh}1: eq.~(\ref{gl:MH}) with the non-conserved noise (a).
\item {\sc mh}2: eq.~(\ref{gl:MH}) with the non-conserved, long-ranged noise (b).
\item {\sc mh}c: eq.~(\ref{gl:MH}) with the conserved noise (c) \cite{Baum07}.
\end{enumerate}
To these, one may add  the spherical model with a conserved order-parameter
(model B dynamics) \cite{Kiss92,Maju95,Sire04,Baum07} which for $d>4$ 
reduces to {\sc mh}c. For $2<d<4$ an explicit, if rather lengthy, expression for 
$C(t,s)$ and its explanation in terms of LSI has been derived \cite{Baum07}. 

In the models {\sc ew}1 and {\sc mh}c the noise is in agreement with detailed
balance while for the other models it is not. Solving the linear equations
(\ref{gl:EW}) and (\ref{gl:MH}) is straightforward and the 
two-time autocorrelations are listed in table~\ref{tb:results_fun}. 
The result quoted for the {\sc mh}c model is only valid 
for $d>2$ as stated; for $d=2$ the scaling function becomes
$f_C(y)=2D \ln[(y-1)/(y+1)]$ \cite{Baum07}. Detailed simulations show that 
the correlation and response functions of the well-known 
Family model \cite{Fami86} and of a variant of it are perfectly
described by the {\sc ew}1 model and hence should be in the same universality
class \cite{Roet06}. 

\subsection{Local scale-invariance for $z=4$} 

In line with the approach taken for $z=2$, we shall consider the dynamical
symmetries of the `deterministic' part of these equations and then try to
prove a generalization of the Bargman superselection rules such that the 
results can be extended to the full noisy system. 
Consider the `Schr\"odinger equation' ${\cal S}_4 \psi=0$ where
the `Schr{\"o}dinger operator' is defined as
\BEQ
\label{mh}
\mathcal{S}_4 := -\mu \partial_t + \frac{1}{16}\partial_r^4
\EEQ
The parameter $\mu$ is the analogue of the mass
$\mathcal{M}$ for the case $z = 2$ (and should not be confused with the 
exponent $\mu$ defined in section~3). Dynamical symmetries of this kind of 
simple linear equation have been constructed for any given $z$ \cite{Henk02}. 
For the special case $z=4$, some of the generators read in $d=1$ dimensions 
as follows (see \cite{Roet06,Baum07} for the extension to $d>1$)
\BEA
\label{generators:xm1}
X_{-1} & = & - \partial_t \nonumber \\
Y_{-1/4} & = & -\partial_r \nonumber \\
X_0 &= & -t \partial_t - \frac{1}{4} r \partial_r -
\frac{x}{4} \nonumber \\
X_1 &= & - t^2 \partial_t - \frac{1}{2} t r \partial_r
- \frac{x}{2} t - \mu r^2 
\partial_r^{-2} + 4 \gamma  r
\partial_r^{-3} - 6 \gamma \partial_r^{-4} \label{generators:y2}\\
Y_{3/4} &=& -t \partial_r - 4 \mu r
\partial_r^{-2} +8 \gamma 
\partial_r^{-3}
\nonumber 
\EEA
where $\gamma$ is a parameter and the `derivative' $\partial_r^a$ satisfies 
the formal properties $\partial_r^a \partial_r^b =\partial_r^{a+b}$ and 
$[\partial_r^a,r] = a \partial_r^{a-1}$ \cite{Henk02}. Then 
it is straightforward to check that 
\BEQ
{} [{\cal S}_4,Y_{-1/4}] \:=\: [{\cal S}_4,Y_{3/4}] 
\:=\: [{\cal S}_4,R^{(i,j)}] \:=\: 0 \;\; , \;\;
{} [{\cal S}_4,X_0] = - {\cal S}_4 
\EEQ
and
\BEQ
[\mathcal{S}_4,X_1] \:=\: -2 t\mathcal{S}_4 + \frac{\mu}{2}\left(x-\frac{3}{2} 
+ \frac{2\gamma}{\mu}\right)
\EEQ
and the commutator of ${\cal S}_4$ with all other generators 
(\ref{generators:xm1}) vanishes. As before for Schr\"odinger and
conformal invariance, this
means the generators (\ref{generators:xm1}) act as dynamical symmetry
operators provided that the scaling dimension of the solution $\psi$
of ${\cal S}_4\psi=0$ is $x = 3/2 - 2 \gamma/\mu$.
As before, a scaling operator is called {\em quasiprimary} 
\cite{Bela84,Henk02,Henk06g} if its infinitesimal
transformation is given by (\ref{generators:xm1}). Quasiprimary operators
are now characterized  by the triplett $(x,\gamma/\mu,\mu)$, rather than the
pair $(x,{\cal M})$ as it is the case for Schr\"odinger-invariance. These
'quantum numbers' are connected to the critical exponents of
the model at hand. It appears that $x$ and $\gamma/\mu$ are universal numbers,
while $\mu$ (as is $\cal M$) is dimensionful and must be non-universal. 

The two-point function $F_0^{(2)}$ built from two quasiprimary scaling operators
$\phi$ and $\wit{\phi}$ has the form \cite{Henk02,Roet06,Baum07}
\BEQ
\left\langle
\phi(t,\vec{r}) \wit{\phi}(s,\vec{r}')\right\rangle_0 = \delta_{x,\wit{x}}\,\delta_{\gamma,\wit{\gamma}}\,\delta_{\mu,\wit{\mu}}\: 
(t-s)^{-x/2} \varphi\left(|\vec{r}-\vec{r}'|(t-s)^{-1/4}\right)
\EEQ
where the scaling function is given by 
$\varphi(u)=\sum_{m  = 0}^3 c_m \varphi^{(m)}(u)$ where
\begin{equation}
\label{result:sf}
\hspace{-1.4cm}
\varphi^{(m)} (u)= \sum_{n = 0}^\infty b_n^{(m)} u^{4 (n-1) +  m +1}
\;\;,\;\; 
b_n^{(m)} = \frac{(-16 \mu)^n \Gamma \left( 1 + m \right)
\Gamma \left(n + \frac{m}{4} -\frac{\gamma}{\mu}
\right)}{\Gamma \left(4 (n-1) +
m +2 \right) \Gamma \left( \frac{m}{4}
-\frac{\gamma}{\mu} \right)}.
\end{equation}
Here, the coefficients $c_m$ are left arbitrary (although further constraints
may be imposed on them by requiring that the scaling function $\vph(u)\to 0$
as $u\to\infty$).

In order to break time-translation invariance as required for applications
to ageing, we introduce a time-dependent potential $v(t)$ 
and consider the equation
\BEQ
\label{spherical}
\partial_t \phi = -\frac{1}{16 \mu}
\nabla_{\vec{r}}^2\Big(-\nabla_{\vec{r}}^2 \phi + v(t) \phi
\Big) + \eta(t,\vec{r}) + j(t,\vec{r})
\EEQ
with a gaussian white noise $\eta(t,\vec{r})$ and a perturbation
$j(t,\vec{r})$ for the calculation of responses. The potential can be
eliminated through a gauge transformation \cite{Baum07}
\BEQ
\phi(t,\vec{r}) = \exp\left(-\frac{1}{16 \mu} \int_0^t \D
\tau v(\tau) \nabla_{\vec{r}}^2 \right) \psi(t,\vec{r})
\EEQ
In contrast to the case $z=2$, we now assume that for $t\to\infty$
$\int_0^t \!\D \tau\, v(\tau) \sim t^\digamma$ 
in order to ensure scaling behaviour for large times. This
defines the new parameter $\digamma$ but we warn the reader that this
parameter should not be confused with the one used in eq.~(\ref{3:kvf}). 

Now, progress is possible if the deterministic part reduces to a {\em linear} 
equation. Then  Wick's theorem holds which permits to derive analogous
reduction formul{\ae} as in the case $z=2$. For 
a fully disordered initial state with 
$\langle \phi(0,\vec{R}) \phi(0,\vec{R}') \rangle = a_0
\delta(\vec{R}-\vec{R}')$ 
the response function is then explicitly given by \cite{Roet06,Baum07}
\BEA 
\hspace{-2.0cm}
R(t,s;\vec{r}) &=& F_0^{(2)}(t,s;\vec{r})
\nonumber \\
\hspace{-2.0cm}&=& 
(t-s)^{-x/2} \sum_{m=0}^3 c_m u^{m+1}
\sum_{k,n = 0}^\infty b_{n,k}^{(m)}\left[-\frac{\kappa_0
(t^\digamma - s^\digamma)}{16 \mu
(t-s)^{1/2}} \right]^k u^{4(n-1) - 2 k}
\label{F02}
\EEA
with the scaling variable $u = |\vec{r}| (t-s)^{-1/4}$ 
and the coefficients $b_{n,k}^{(m)}$, which are given by
\BEQ
b_{n,k}^{(m)} = \frac{1}{k!} 
\frac{\Gamma(4(n-1) + m +2)}{\Gamma(2 k - 4(n-1) -m)} \, b_n^{(m)}
\EEQ
which is perfectly consistent with the available information from the
several {\sc mh} models introduced above \cite{Roet06} and also for the 
spherical model with a conserved order-parameter \cite{Baum07}.

On the other hand, the correlation function becomes, with $t>s$
\BEA
C(t,s;\vec{r}) &=& \frac{a_0}{2} \int_{\mathbb{R}^d} 
\!\D \vec{R}\, F_0^{(3)}(t,s,0;
\vec{r},\vec{0},\vec{R})\nonumber \\ 
& & -\frac{T}{16\mu} \int_0^s \!\D\tau \int_{\mathbb{R}^d}\!\D\vec{R} \,
F_0^{(3)}(t,s,\tau;\vec{r},\vec{0},\vec{R})
\EEA
where the deterministic three-point function is \\
$F_0^{(3)}(t,s,\tau;\vec{r},\vec{r}',\vec{R}) := 
\langle \phi(t,\vec{r})\phi(s,\vec{r}') \wit{\phi}^2(\tau,\vec{R}) \rangle_0$.
If the deterministic part reduces to
a linear equation, then Wick's theorem leads to \cite{Baum07}
\BEQ
F_0^{(3)}(t_1,t_2,t_3; \vec{r}_1,\vec{r}_2,\vec{r}_3) = 2
F_0^{(2)}(t_1,t_3;\vec{r}_1-\vec{r}_3) F_0^{(2)}(t_2,t_3;\vec{r}_2-\vec{r}_3)
\EEQ
where $F_0^{(2)}$ can be taken from eq.~(\ref{F02}). 
Carrying out the (rather long) calculation then allows to fully reproduce 
all results listed in table~\ref{tb:results_fun} for the models with $z=4$. 
At least for the models studied so far with $z=4$, 
the relation $a=a'$ was empirically found to be satisfied. That is not
surprising in view  of the simple linear equation satisfied by the 
order-parameter field $\phi$.

The {\sc mh} models considered in this section are, together with the
critical spherical model with a conserved order-parameter, 
the first analytically 
solved examples with $z\ne 2$ where local scale-invariance could be fully 
confirmed. These examples make it in particular clear that the height of the
surface in growth processes is a natural candidate for being described by a 
quasiprimary scaling operator of local scale-invariance.

\subsection{`Conformal invariance' for $z\ne 1,2$ ?}

We finish with a short speculation on how one might construct an analogy of the
extension $\mathfrak{sch}_d \to \mathfrak{conf}_{d+2}$ discussed in 
section~2.3. By analogy, one might  try to consider the `mass' $\mu$ as a
further variable and thus try to find 
analogues to the operators $V_{-}$ and $N$ described in section~2.3.
If we very na\"{\i}vely generalise $V_{-}$ and $N$
from the case $z = 2$ and define
\BEQ
V_{-} = - \II \partial_\mu \partial_r^{3} + 4 \II r \partial_t \;\; , \;\;
N  =  - t\partial_t - 1 - \mu \partial_\mu
\EEQ
then it can be checked that both commutators
$[{\cal S}_4,N]$ and $[{\cal S}_4,V_{-}]$  vanish.
This already implies that these two operators describe new,
non-trivial symmetries and are appropriate candidates for
a possible extension of the algebra freely generated from the minimal set 
$\{ X_{-1},X_0,X_1,Y_{-1/4},Y_{3/4}\}$.
At this stage, the algebraic structure of such an extension is not clear at 
all; see \cite{Henk02} for the discussion of some other difficulties which
may arise when one wishes  to find the algebra  satisfied  by local 
scale-transformations with $z\ne 1,2$ and such that they can be symmetries
of physically non-trivial equations -- which in the context of ${\cal S}_4$ 
would mean that $\mu \ne 0,\infty$. 

Of course, the central question whether an analogue of 
Bargman superselection rules exists is not even addressed. We hope to come
back to this elsewhere, see \cite{Baum07b}.

\section{Conclusions}

We have considered how local scale-invariance might be used in order to
arrive at an explicit prediction for the two-time autocorrelation function
of phase-ordering kinetics (as well as for non-equilibrium critical dynamics in those special cases where $z=2$) capable to reproduce simulational data. 
In comparison with the two-time response function, whose functional form
simply follows from the assumed covariance of the quasiprimary scaling
operators of the ageing algebra $\mathfrak{age}_d$, the calculation of
$C(t,s)$ does require conceptually important extensions. First, the explicit
reduction from a stochastic Langevin equation to a deterministic equation is
needed since a simple covariance argument for $C(t,s)$ would have led to
a vanishing autocorrelation because of the Bargman superselection rules, see
\cite{Pico04}. 
Second, invariance under $\mathfrak{age}_d$ alone does not seem to lead to
any predictions for the autocorrelator, 
with the exception of the exponent relation 
$\lambda_C=\lambda_R$ between the autocorrelation and the autoresponse exponents 
for the case of fully disordered initial conditions. Therefore, the third
required extension concerns the dynamical symmetry algebra
itself 
\BEQ
\mathfrak{age}_d \to \mathfrak{sch}_d \to \mathfrak{conf}_{d+2}
\EEQ
which then allows to reduce the scaling function $f_C(y)$ to the linear
combination of two known functions. We have also described some subleties in the
precise definition of quasiprimary scaling operators and the several
scaling dimensions ($x,\xi,\digamma$) on which they depend. 

Besides several confirmations of the first two steps from analytically solved
models with an underlying free-field theory, remarkably there is strong
evidence from two-dimensional kinetic Ising and Potts models in favour
of the last extension to a new type of conformal invariance, at least for 
quenches to $T<T_c$, where $z=2$.

It might be useful here to list some of the important open problems. 
\begin{enumerate}
\item A more systematic approach replacing the glueing procedure outlined
in section~3.2 (and the appendix) should be found, 
hopefully leading to full agreement between the theory and the data. 
\item The Galilei- and other local scale-invariances of {\em non-linear}
partial differential equations must be studied. Present results \cite{Stoi05,Baum05b} require Galilei-invariance for all times which is too
strong a requirement to be able to include the usual kind of Langevin equations
\cite{Hohe77}. A new kind of asymptotic Galilei-invariance, only valid
in the scaling regime (\ref{1:validite}), 
would be enough and might be more flexible, but remains to be constructed. 

The importance of Galilei-invariance implies that models without a spatial
structure such as the zero-range process \cite{Evan05,Godr06} cannot be expected
to satisfy a local scale-invariance. On the other hand, it might be of interest
to inquire whether driven diffusive systems \cite{Schm95} may possess 
some form of local scale-invariance. 
\item Analytical tests of local scale-invariance only exist for a rather
simple kind of solvable systems, with an underlying free-field theory. The
calculation of two-time observables in more rich integrable systems would 
certainly provide most useful information. 
\item Present tests of local scale-invariance have been limited mainly
to systems with $z=2$ or focused on the autoresponse whose form does not depend
on directly on $z$. Therefore important aspects of the theory as outlined
in \cite{Henk02} have not yet been tested. We hope that the consideration
of phase-ordering in disordered Ising models \cite{Paul04,Henk06b} and in
certain models with long-range interactions \cite{Cann01,Baum07a} 
where $z$ depends
continuously on the control parameters of the model, will allow to do this. 
\item A central point for the applicability to stochastic Langevin equations
is the derivation of extensions of the Bargman superselection rules to $z\ne 2$. 
Work along these lines is in progress \cite{Baum07b}. 
\item Is there a way to formulate an infinite-dimensional extension of the
local scale-transformations considered here, in analogy with $2D$ conformal
invariance {\em at} equilibrium \cite{Bela84,Card90}? At the time of writing, 
there merely exist some isolated mathematical results on equations with an infinite-dimensional dynamical symmetry, on the algebraic structure of 
the related Lie algebras and their vertex operator representations 
\cite{Cher04,Roge06,Unte07}. What would be the physical signatures of such
a symmetry which could be tested in specific models~?
\item Dynamical scaling combined with reparametrization invariance 
has been considered for glasses
where, because of logarithmic scaling, the effective dynamical exponent
$z\to\infty$, see \cite{Cham02,Cham05} and references therein. Will there be a 
way to relate these distinct approaches of local scaling~? 
\end{enumerate} 
All in all, local scale-invariance may offer an approach complementary to 
numerical simulation and the field-theoretical renormaliztion group (so far
restricted to non-equilibrium {\em critical} dynamics \cite{Cala05}). The
presently available evidence suggests the possibility of establishing a 
hidden dynamical symmetry which had not been suspected and, if proven correct, 
could simplify considerably the 
description of the non-equilibrium dynamics of many-body systems. 

\annexe{}

We explain how to extend the description of the three-point function
$\langle\Phi(t)\Phi(s)\wit{\Phi}^2(u)\rangle_0$ to the region where $t\approx s$
and local scale-invariance can no longer be used. Throughout, we restrict to 
the case of phase-ordering ($T<T_c$) and we shall derive 
eq.~(\ref{gl:PSIFinal}). Our physical criterion is
the requirement that the two-time autocorrelator be symmetric in $t$ and $s$ and
furthermore be non-singular as $t-s\to 0$. From the from eq.~(\ref{glC3}) 
this leads to the following form of the scaling function $\Psi(\rho)$,
since $x-\wit{x}_2+2\xi-2\wit{\xi}_2=\lambda_C-d/2$
\BEQ \label{gl:A:1}
\Psi(\rho) = \rho^{\lambda_C-d/2} \sum_{n=0}^{\infty} \Psi_n \rho^{2n}
\EEQ
where the $\Psi_n$ are constants. Because of the Yeung-Rao-Desai inequality
$\lambda_C\geq d/2$ \cite{Yeun96} this expression will increase with $\rho$ for
sufficiently small values of $\rho$,
whereas the LSI-prediction eq.~(\ref{gl:3:Psipo}) gives a divergence 
at $\rho=0$, unless $a'\leq 0$ which in any case is incompatible with the
explicit results in many models, see section~4 and \cite{Henk07a,Henk07b}. 
We expect (\ref{gl:A:1}) to be valid for sufficiently small values of $\rho$
and to change to the form predicted by LSI if $\rho$ becomes larger than some
minimal value $\eps$. For $\rho$ sufficiently small, we may approximate
(\ref{gl:3:Psipo}) as follows
\BEA \label{gl:A:2}
\Psi_{\rm scal}(\rho) &\simeq& \psi_0 \rho^{\lambda_C-d/2-2a'}
\left(1-\frac{2\lambda_C-d-2a'}{1+\lambda_C-d/2-2a'}{\cal M}\rho +\ldots\right)
\nonumber \\
& & +\psi_1 {\cal M}^{2a'+d/2-\lambda_C} +\ldots
\EEA
which should be valid for $\rho\gtrsim \eps$. Since in phase-ordering one
usually treats systems of class S where $a=a'=1/z=1/2$ and $\lambda_C\geq d/2$, 
all the terms retained here are 
needed to have a consistent expansion as $\rho\to 0$. 

If the cross-over point $\eps$ is sufficiently small, only the leading term
in (\ref{gl:A:1}) is needed. We construct a scaling function $\Psi(\rho)$ 
for all $\rho\geq 0$ by requiring that the branches (\ref{gl:A:1}) and
(\ref{gl:A:2}) meet  continuously at $\rho=\eps$. This condition 
fixes $\Psi_0=\Psi_0(\eps)$ and we find, to this order
\BEA
\Psi_0(\eps) &=& \psi_0 \eps^{-2a'} -\psi_0{\cal M}
\frac{2\lambda_C-d-2a'}{1+\lambda_C-d/2-2a'}\eps^{1-2a'}
\nonumber \\
& & + \psi_1 {\cal M}^{2a'+d/2-\lambda_C} \eps^{d/2-\lambda_C}
\label{gl:A:3}
\EEA

The required scaling function $\vec{\Psi}$ is given by
\BEA
\vec{\Psi}(w) &=& \int_{\mathbb{R}^d} \!\D \vec{R}\: 
e^{-{\cal M} w \vec{R}^2 /2}\, \Psi\left(\frac{\vec{R}^2}{2}\right)
\nonumber \\
&=& 2^{d/2-1} S_d \int_0^{\infty} \!\D u\: u^{d/2-1}\, e^{-{\cal M}w u}\,
\Psi(u) 
\nonumber \\
&=& 2^{d/2-1} S_d \left( \int_0^{\infty} \!\D u\: u^{d/2-1}\, e^{-{\cal M}w u}
\,\Psi_{\rm LSI}(u) \right.
\nonumber \\
& & + \left. \int_0^{\eps} \!\D u\: u^{d/2-1}\, e^{-{\cal M}w u}
\left( \Psi_0(\eps) u^{\lambda_C-d/2} -\Psi_{\rm scal}(u)\right) \right)
\nonumber \\
&=:& 2^{d/2-1} S_d \left(  I_0 +  I_1 + I_{\eps} \right) 
\EEA
where $S_d$ is the surface of the sphere in $d$ dimensions, $I_{0,1}$ are proportional to $\psi_{0,1}$, $\Psi_{\rm LSI}$
is the LSI-prediction given in (\ref{gl:3:Psipo}) and eqs.~(\ref{gl:A:2}) and
(\ref{gl:A:3}) were used. In the third line the range of integration was
split into the intervals $[0,\eps]$ and $[\eps,\infty]$ and the leading term
from (\ref{gl:A:1}) was used in the first subinterval. The terms $I_{0,1}$
refer to the two independent solutions for $\Psi(\rho)$ in 
eq.~(\ref{gl:3:Psipo}). They can be calculated using the identity
\BEQ
\int_0^{\infty}\!\D u\: u^{\alpha-1}\, e^{-\omega u}\: {_1F_1}(A,B;-u)
=\frac{\Gamma(\alpha)}{\omega^{\alpha}} \,
{_2F_1}\left(A,\alpha;B;-\frac{1}{\omega}\right)
\EEQ
This already gives the first two lines in (\ref{gl:PSIFinal}). 
The terms contained in $I_{\eps}$ are calculated from the identity
\BEQ
\int_0^1 \!\D u\: u^{\beta-1}\, e^{-\omega u} 
= \omega^{-\beta}\, \gamma(\beta,\omega)
\EEQ
where $\gamma(\beta,\omega)$ is an incomplete gamma function \cite{Abra65}. 
Combining the various terms and setting
\BEQ
A := \psi_0 2^{d/2-1} S_d {\cal M}^{2a'-\lambda_C} \;\;,\;\;
B := \psi_1 2^{d/2-1} S_d {\cal M}^{2a'-\lambda_C} \;\;,\;\;
E := {\cal M}\eps
\EEQ
we finally arrive at eq.~(\ref{gl:PSIFinal}) in the text. 

Inserting this result into the two-time autocorrelation function 
eq.~(\ref{3:fCpo}), the amplitude $C_{\infty}=\vec{\Psi}(1)=\lim_{y\to\infty}
y^{\lambda_C/2} C(ys,s)$ is given by 
\BEA
\hspace{-1.5truecm}C_{\infty} &=& A \left\{ \Gamma(\lambda_C-2a')\: {_2F_1}(2\lambda_C-d-2a',
\lambda_C-2a';1+\lambda_C-d/2-2a';-1) \right.
\nonumber \\
\hspace{-1.5truecm}& & + E^{-2a'}\gamma(\lambda_C,E)-\gamma(\lambda_C-2a',E) \nonumber \\
\hspace{-1.5truecm}& & + \left. E^{1-2a'}\frac{2\lambda_C-d-2a'}{1+\lambda_C-d/2-2a'}
\left[E^{2a'-1}\gamma(\lambda_C+1-2a',E)-\gamma(\lambda_C,E)\right] \right\}
\nonumber \\
\hspace{-1.5truecm}&+& B\left\{ \Gamma(d/2)\:{_2F_1}(\lambda_C-d/2,d/2;1+2a'+d/2-\lambda_C;-1)
\right. 
\nonumber \\
\hspace{-1.5truecm}& & \left. +E^{d/2-\lambda_C} 
\gamma(\lambda_C,E)-\gamma(d/2,E) \right\}
\EEA
which gives a useful technique to fix one of the free parameters $A,B,E$ 
in terms of the other two. Indeed, in order to find $C_{\infty}$ empirically
from numerical data, it is considerably more precise to plot 
$C(ys,s) y^{-\lambda_C/2} (y-1)^{\lambda_C}$ over against $y$, rather than
the simplistic  $C(ys,s) y^{\lambda_C/2}$. 

The opposite limit $\lim_{y\to 1} C(ys,s) = C(t,t) = C_1$ exists if 
$2a'\leq 1$ and then becomes
\BEQ
\hspace{-2.2truecm}
C_1 = \Gamma(\lambda_C)2^{-\lambda_C} \left\{ A E^{-2a'} \left[
1 -E\frac{2\lambda_C-d-2a'}{1+\lambda_C-d/2-2a'}\left(1-\delta_{2a',1}
\right)\right] 
+ B E^{d/2-\lambda_C} \right\} ~~
\EEQ
which for finite $E$ is indeed a finite constant. 

For a free field, one has $\lambda_C=d/2$. If one takes $A=0$ and $B>0$, one
recovers $\vec{\Psi}(w)=B\Gamma(d/2) w^{-d/2}$, as it should be.

If one should find that
the description of $\Psi(\rho)$ for $\rho\leq \eps$ 
merely by the lowest order-term is not
sufficiently precise, further terms from (\ref{gl:A:1}) may be included 
(balanced by the higher-order terms of the expansion of $\Psi_{\rm LSI}(\rho)$) 
and the coefficients $\Psi_n$ fixed by requiring
also the continuity of the respective derivatives of $\Psi(\rho)$ at
$\rho=\eps$.

\zeile{1}

\noindent 
{\large\bf Acknowledgements:} 

It is a pleasure to thank the organisers of the workshop {\it Principles of 
the dynamics of non-equilibrium systems} and 
Isaac Newton Institute for warm hospitality in a stimulating environnement; 
A.J. Bray, L. Berthier, J.L. Cardy, J.P. Garrahan, A. Lef\`evre, 
S. Majumdar, G. Odor, V. Rittenberg, B. Schmittmann, G.M. Sch\"utz, P. Sollich for fruitful discussions; 
A. Picone, M. Pleimling, A. R\"othlein, S. Stoimenov, J. Unterberger 
for the year-long collaborations which lead to the results 
reviewed here and E. Lorenz and W. Janke for communicating their
results before publication, discussions and for sending figure~\ref{fig:C(t,s)}. 
We also thank A. Gambassi and H. Hinrichsen for their critical but 
constructive comments on LSI,  forcing us to lay the foundations of the 
theory in a more precise way. 
FB acknowledges the support by the Deutsche Forschungsgemeinschaft through grant 
no. PL 323/2. This work was also suppoted by the franco-german binational PAI 
programme PROCOPE. 

\newpage




\begin{thebibliography}{999}
\bibitem{Abra65} M. Abramowitz and I.A. Stegun, {\it Handbook of mathematical
functions}, Dover (New York  1965). 
\bibitem{Anni06} A. Annibale and P. Sollich, J. Phys. {\bf A39}, 2853 (2006).
\bibitem{Barg54} V. Bargman, Ann. of Math. {\bf 56}, 1 (1954).
\bibitem{Baru73} A.O.  Barut, Helv. Phys. Acta {\bf 46}, 496 (1973). 
\bibitem{Baum05a} F. Baumann, M. Henkel, M. Pleimling and J. Richert, 
J. Phys. A: Math. Gen. {\bf 38} 6623, (2005).     
\bibitem{Baum05b} F. Baumann, S. Stoimenov and M. Henkel, J. Phys. {\bf A39}, 
4095 (2006). 
\bibitem{Baum06a} F. Baumann and M. Pleimling, J. Phys. {\bf A39}, 1981 (2006).
\bibitem{Baum06d} F. Baumann and A. Gambassi, J. Stat. Mech. P01002 (2007)
\bibitem{Baum07} F. Baumann and M. Henkel, J. Stat. Mech. P01012 (2007).
\bibitem{Baum07a} F. Baumann, S.B. Dutta and M. Henkel, {\tt en pr\'eparation}. 
\bibitem{Baum07b} F. Baumann and M. Henkel, {\tt en pr\'eparation}. 
\bibitem{Bela84} A.A. Belavin, A.M. Polyakov and A.B. Zamolodchikov, Nucl.
Phys. {\bf B241}, 333 (1984). 
\bibitem{Bert99} L. Berthier, J.L. Barrat and J. Kurchan, Eur. Phys. J.
{\bf B11}, 635 (1999).
\bibitem{Bert01} L. Berthier, P.C.W. Holdsworth and M. Sellitto, J. Phys.
{\bf A34}, 1805 (2001). 
\bibitem{Boye76} C.D. Boyer, R.T. Sharp and P. Winternitz, J. Math. Phys. 
{\bf 17}, 1439 (1976).
\bibitem{Bray91} A.J. Bray and S. Puri, Phys. Rev. Lett. {\bf 67}, 2670 (1991).
\bibitem{Bray94} A.J. Bray, Adv. Phys. {\bf 43}, 357 (1994). 
\bibitem{Bray94b} A.J. Bray and A.D. Rutenberg, Phys. Rev. {\bf E49}, 
R27 (1994); {\bf E51}, 5499 (1995). 
\bibitem{Brow97} G. Brown, P.A. Rikvold, M. Suton and M. Grant, 
Phys. Rev. {\bf E56}, 6601 (1997).
\bibitem{Burd73} G. Burdet, M. Perrin and P. Sorba, Comm. Math. Phys. 
{\bf 34}, 85 (1973). 
\bibitem{Cala05} P. Calabrese and A. Gambassi, J. Phys. {\bf A38}, R181 (2005).
\bibitem{Cala06} P. Calabrese, A. Gambassi and F. Krzakala, 
J. Stat Mech. Theor. Exp. P06016 (2006). 
\bibitem{Cala07} P. Calabrese and A. Gambassi, J. Stat. Mech. P01001 (2007). 
\bibitem{Cann01} S.A. Cannas, D.A. Stariolo and F.A. Tamarit, Physica 
{\bf A294}, 362 (2001). 
\bibitem{Card90} J.L. Cardy in E. Br\'ezin and J. Zinn-Justin (eds), 
{\it Fields, strings and critical phenomena}, Les Houches XLIX, North
Holland  (Amsterdam 1990).
\bibitem{Cate00} M.E. Cates and M.R. Evans (eds)
{\it Soft and fragile matter}, IOP Press (Bristol 2000).
\bibitem{Cham02} C. Chamon, M.P. Kennett, H. Castillo and L.F. Cugliandolo,
Phys. Rev. Lett. {\bf 89}, 217201 (2002). 
\bibitem{Cham05} C. Chamon, L.F. Cugliandolo and H. Yoshino, J. Stat. Mech. 
Theory Exp. P01006 (2006). 
\bibitem{Cher04} R. Cherniha and M. Henkel, J. Math. Anal. Appl. {\bf 298}, 
487 (2004). 
\bibitem{Cris03} A. Crisanti and F. Ritort, J. Phys. {\bf A36}, R181 (2003)
\bibitem{Cugl02} L.F. Cugliandolo, in {\it Slow Relaxation and
non equilibrium dynamics in condensed matter}, Les Houches Session 77 July 2002,
J-L Barrat, J Dalibard, J Kurchan, M V Feigel'man eds (Springer, 2003); 
also available at {\tt cond-mat/0210312}.
\bibitem{deDo78} C. de Dominicis and L. Peliti, Phys. Rev. {\bf B18}, 
353 (1978).
\bibitem{deOl93} M.J. de Oliveira, J.F.F. Mendes and M.A. Santos, 
J. Phys. {\bf A26}, 2317 (1993).
\bibitem{Dorn98} I. Dornic, th\`ese de docotorat, Nice et Saclay 1998.
\bibitem{Edwa82} S.F. Edwards and D.R. Wilkinson, Proc. Roy. Soc. London Ser. 
{\bf A381}, 17 (1982). 
\bibitem{Evan05} M.R. Evans and T. Hanney, J. Phys. {\bf A38}, R195 (2005).
\bibitem{Fami86} F. Family, J. Phys. {\bf A19}, L441 (1986).  
\bibitem{Fedo06} A.A. Fedorenko and S. Trimper, Europhys. Lett. {\bf 74},
89 (2006). 
\bibitem{Fish88} D.S. Fisher and D.A. Huse, Phys. Rev. {\bf B38}, 373 (1988).
\bibitem{Fush93} W.I. Fushchich, W.M. Shtelen and N.I. Serov, {\it Symmetry analysis and exact solutions of equations of nonlinear mathematical physics},
Kluwer (Dordrecht 1993). 
\bibitem{Gamb06} A. Gambassi, J. Phys. Conf. Series {\bf 40}, 13 (2006). 
\bibitem{Giul96} D. Giulini, Ann. of Phys. {\bf 249}, 222 (1996). 
\bibitem{Glau63} R.J. Glauber, J. Math. Phys. {\bf 4}, 294 (1963). 
\bibitem{Godr00a} C. Godr\`eche and J.-M. Luck, J. Phys. {\bf A33}, 1151 (2000).
\bibitem{Godr00b} C. Godr\`eche and J.-M. Luck, J. Phys. {\bf A33}, 9141 (2000).
\bibitem{Godr02} C. Godr\`eche and J.M. Luck, J. Phys. Cond. Matt. {\bf 14},
1589 (2002).
\bibitem{Godr06} C. Godr\`eche, in \cite{Henk06} ({\tt cond-mat/0604276}). 
\bibitem{Hage72} C.R. Hagen, Phys. Rev. {\bf D5}, 377 (1972). 
\bibitem{Henk94} M. Henkel, J. Stat. Phys. {\bf 75}, 1023 (1994). 
\bibitem{Henk02} M. Henkel, Nucl. Phys. {\bf B641}, 405 (2002).
\bibitem{Henk02a} M. Henkel, M. Paessens and M. Pleimling, 
Europhys. Lett. {\bf 62}, 644 (2003)
\bibitem{Henk03} M. Henkel and J. Unterberger, Nucl. Phys. {\bf B660}, 
407 (2003). 
\bibitem{Henk03d} M. Henkel and G.M. Sch\"utz, J. Phys. {\bf A37}, 591 (2004).
\bibitem{Henk03e} M. Henkel, M. Paessens and M. Pleimling,
Phys. Rev. {\bf E69}, 056109 (2004).
\bibitem{Henk04b} M. Henkel, A. Picone and M. Pleimling, Europhys. Lett. 
{\bf 68}, 191 (2004).   
\bibitem{Henk06} M. Henkel, M.  Pleimling and R. Sanctuary (eds), {\it
Ageing and the glass transition}, Springer Lecture Notes in Physics {\bf 716}, 
Springer (Heidelberg  2007).  
\bibitem{Henk06a} M. Henkel, T. Enss and M. Pleimling, J. Phys. {\bf A39}, L589
(2006). 
\bibitem{Henk06b} M. Henkel and M. Pleimling, Europhys. Lett. {\bf 76}, 561
(2006). 
\bibitem{Henk06g} M. Henkel and J. Unterberger, Nucl. Phys. 
{\bf B746}, 155 (2006). 
\bibitem{Henk06h} M. Henkel, R. Schott, S. Stoimenov and J. Unterberger,
submitted to Quantum Probability, {\tt math-ph/0601028}). 
\bibitem{Henk07a} M. Henkel, J. Phys. Cond. Matt. {\bf 19}, 065101 (2007).
\bibitem{Henk07b} M. Henkel and M. Pleimling, dans W. Janke (ed)
{\it Rugged free-energy landscapes: common computational approaches 
in spin glasses, structural glasses and biological macromolecules}, 
Springer Lecture Notes in Physics, Springer (Heidelberg 2007).  
\bibitem{Hohe77} P. Hohenberg and B.I. Halperin, Rev. Mod. Phys. {\bf 49}, 
435 (1977). 
\bibitem{Houc02} B. Houchmandzadeh, Phys. Rev. {\bf E66}, 052902 (2002). 
\bibitem{Howa97} M. Howard and U.C. T\"auber,  J. Phys. {\bf A30}, 7721 (1997).
\bibitem{Huse89} D.A. Huse, Phys. Rev. {\bf B40}, 304 (1989). 
\bibitem{Jank06} W. Janke, in \cite{Henk06}. 
\bibitem{Jans89} H.K. {\cal J}anssen, B. Schaub and B. Schmittmann, Z. Phys. {\bf B73}, 539 (1989). 
\bibitem{Jans92} H.K. {\cal J}anssen, in G. Gy\"orgyi et {\em al.} (eds) {\it
{}From Phase transitions to Chaos}, World Scientific (Singapour 1992), p. 68
\bibitem{Kast68} H.A. Kastrup, Nucl. Phys. {\bf B7}, 545 (1968).
\bibitem{Kiss92} J.G. Kissner, Phys. Rev. {\bf B46}, 2676 (1992). 
\bibitem{Knap86} A.W. Knapp, {\it Representation theory of semisimple groups: an
overview based on examples}, Princeton University Press (Princeton 1986). 
\bibitem{Lipp00} E. Lippiello and M. Zannetti, Phys. Rev. {\bf E61}, 
3369 (2000).
\bibitem{Liu91} F. Liu and G.F. Mazenko, Phys. Rev. {\bf B44}, 9185 (1991).
\bibitem{Lore06} E. Lorenz, {\it Ageing phenomena in phase-odering kinetics in Potts models}, Diplomarbeit Leipzig 31$^{\rm st}$ of July 2005. 
\bibitem{Lore07} E. Lorenz and W. Janke, Europhys. Lett. {\bf 77}, 10003 (2007). 
\bibitem{Maju95} S.N. Majumdar and D.A. Huse, Phys. Rev. {\bf E52}, 270 (1995).
\bibitem{Maze98} G.F. Mazenko, Phys. Rev. {\bf E58}, 1543 (1998).
\bibitem{Maze04} G.F. Mazenko, Phys. Rev. {\bf E69}, 016114 (2004).
\bibitem{Mull63} W. W. Mullins, in {\sl Metal Surfaces:
Structure, Energetics and Kinetics} (Am. Soc. Metal, Metals
Park, Ohio, 1963).
\bibitem{Newm90} T.J. Newman and A.J. Bray, J. Phys. {\bf A23}, 4491 (1990). 
\bibitem{Nied72} U. Niederer, Helv. Phys. Acta {\bf 45}, 802 (1972).
\bibitem{Paes04} M. Paessens and G.M. Sch{\"u}tz, J. Phys. A: Math. Gen. 
{\bf 37}, 4709 (2004). 
\bibitem{Paul04} R. Paul, S. Puri and H. Rieger, Europhys. Lett. {\bf 68}, 
881 (2004).
\bibitem{Perr77} M. Perroud, Helv. Phys. Acta {\bf 50}, 233 (1977).  
\bibitem{Pico02} A. Picone and M. Henkel, J. Phys. {\bf A35}, 5575 (2002).
\bibitem{Pico04} A. Picone and M. Henkel, Nucl. Phys. {\bf B688}, 217 (2004).
\bibitem{Poly70} A.M. Polyakov, Sov. Phys. JETP Lett. {\bf 12}, 381 (1970). 
\bibitem{Roge06} C. Roger and J. Unterberger, Ann. Inst. H. Poincar\'e {\bf 7}, 
1477 (2006) (also available at {\tt math-ph/0601060}). 
\bibitem{Roet06} A. R\"othlein, F. Baumann and M. Pleimling, Phys. Rev. 
{\bf E74}, 061604 (2006). 
\bibitem{Roja99} F. Rojas and A.D. Rutenberg, Phys. Rev. {\bf E60}, 212 (1999).
\bibitem{Sast03} F. Sastre, I. Dornic and H. Chat\'e, Phys. Rev. Lett. {\bf 91},
267205 (2003). 
\bibitem{Schm95} B. Schmittmann and R.K.P. Zia, in C. Domb and J. Lebowitz (eds)
{\it Phase transitions and critical phenomena}, Vol. 17, London (Academic 1995).
\bibitem{Sche03} G. Schehr and P. Le Doussal, Phys. Rev. {\bf E68}, 
046101 (2003). 
27 (2006). 
\bibitem{Sire04} C. Sire, Phys. Rev. Lett. {\bf 93}, 130602 (2004).
\bibitem{Stoi05} S. Stoimenov and M. Henkel, Nucl. Phys. {\bf B723}, 205 (2005).
\bibitem{Stru78} L.C.E. Struik, {\it Physical ageing in amorphous polymers and
other materials}, Elsevier (Amsterdam 1978).
\bibitem{Taeu05} U.C. T\"auber, M. Howard and B.P. Vollmayr-Lee, 
J. Phys. A: Math. Gen. {\bf 38}, R79 (2005).
\bibitem{Taeu06} U.C. T\"auber, in \cite{Henk06} ({\tt cond-mat/0511743}).
\bibitem{Toyo92} H. Toyoki, Phys. Rev. {\bf B45}, 1965 (1992).
\bibitem{Unte07} J. Unterberger, soumis \`a Comm. Math. Phys. 
\bibitem{Wolf90} D. E. Wolf and J. Villain. Europhys. Lett.
{\bf 13}, 389 (1990).
\bibitem{Yeun96} C. Yeung, M. Rao and R.C. Desai, Phys. Rev. {\bf E53}, 3073 (1996). 
\bibitem{Zipp00} W. Zippold, R. K\"uhn and H. Horner, Eur. Phys. J. {\bf B13}, 
531 (2000). 






\end{thebibliography}
\end{document}